\documentclass[fleqn,10pt]{wlscirep}

\usepackage{multirow}
\usepackage{mathrsfs}

\title{High-resolution transport-of-intensity quantitative phase microscopy with annular illumination}

\author[1,2,*]{Chao Zuo}
\author[1,2]{Jiasong Sun}
\author[1,2]{Jiaji Li}
\author[1,2]{Jialin Zhang}
\author[3]{Anand Asundi}
\author[2,*]{Qian Chen}
\affil[1]{Smart Computational Imaging (SCI) Laboratory, Nanjing University of Science and Technology, Nanjing, Jiangsu Province 210094, China}
\affil[2]{Jiangsu Key Laboratory of Spectral Imaging $\&$ Intelligent Sense, Nanjing University of Science and Technology, Nanjing, Jiangsu Province 210094, China}
\affil[3]{Centre for Optical and Laser Engineering (COLE), School of Mechanical and Aerospace Engineering, Nanyang Technological University, Singapore 639798, Singapore}

\affil[*]{Correspondence and requests for materials should be addressed to C.Z. (email: zuochao@njust.edu.cn) or Q.C. (email: chenqian@njust.edu.cn)}

\begin{abstract}
For quantitative phase imaging (QPI) based on transport-of-intensity equation (TIE), partially coherent illumination provides speckle-free imaging, compatibility with brightfield microscopy, and transverse resolution beyond coherent diffraction limit. Unfortunately, in a conventional microscope with circular illumination aperture, partial coherence tends to diminish the phase contrast, exacerbating the inherent noise-to-resolution tradeoff in TIE imaging, resulting in strong low-frequency artifacts and compromised imaging resolution. Here, we demonstrate how these issues can be effectively addressed by replacing the conventional circular illumination aperture with an annular one. The matched annular illumination not only strongly boosts the phase contrast for low spatial frequencies, but significantly improves the practical imaging resolution to near the incoherent diffraction limit. By incorporating high-numerical aperture (NA) illumination as well as high-NA objective, it is shown, for the first time, that TIE phase imaging can achieve a transverse resolution up to 208 nm, corresponding to an effective NA of 2.66. Time-lapse imaging of \emph{in vitro} Hela cells revealing cellular morphology and subcellular dynamics during cells mitosis and apoptosis is exemplified. Given its capability for high-resolution QPI as well as the compatibility with widely available brightfield microscopy hardware, the proposed approach is expected to be adopted by the wider biology and medicine community.
\end{abstract}
\begin{document}

\flushbottom
\maketitle
%
%
\thispagestyle{empty}

\section*{Introduction}

\noindent
In the field of optical microscopy, there has been a continued need towards increasing imaging resolution for visualizing subcellular features of the biological samples. Such a need has driven the development of many unique sub-diffraction imaging techniques that have made great impacts on biological fluorescence imaging, allowing visualization of sample features well beyond the diffraction limit by either single-molecule localization, such as in stochastic optical reconstruction microscopy (STORM) and photo activated localization microscopy (PALM) \cite{1}, or spatially modulated excitation, such as in stimulated emission depletion (STED) \cite{2} and structured illumination microscopy (SIM) \cite{3}. However, such sub-diffraction imaging techniques require fluorescent dyes and fluorescent proteins as biomarkers, and are thus ill-suited for samples that are non-fluorescent or cannot be easily fluorescently tagged. Besides, the photobleaching and phototoxicity of the fluorescent agents prevent live cells imaging over extended periods of time \cite{4}.

During recent years, quantitative phase imaging (QPI) \cite{5,6,7} has emerged as an invaluable optical tool for biomedical research thanks to its unique capabilities to image optical thickness variation of living cells and tissues without the need for specific staining or exogenous contrast agents (\emph{e.g.}, dyes or fluorophores). Quantitative phase profiles of cells allow determination of cellular structure and biophysical parameters with minimal sample manipulation. Especially in cases where conventional preparation techniques, such as fixation, staining, or fluorescent tagging, may affect cellular functions and limit biological insight, QPI offers an important alternative \cite{8,9}. By introducing the principles of interferometry and holography into microscopy, both amplitude and phase information of microscopic specimens can be simultaneously demodulated from the recorded interference patterns \cite{6,10,11,12}. Nevertheless, due to the laser illumination sources typically used, coherent QPI methods suffered from speckle noise arising from stray interferences from imperfections in the optical system. This limitation was overcome by the low-coherence QPI approaches, such as spatial light interference microscopy (SLIM) \cite{13}, white-light diffraction phase microcopy (wDPM) \cite{14}, quadriwave lateral shearing interferometry (QWLSI) \cite{15}, $\tau$ interferometry \cite{16}, \emph{etc}. The combination of broadband illumination with the common-path geometries significantly alleviates the coherent noise problem and enhances the stability to mechanical vibrations and air fluctuations that typically affect any interferometric systems. However, these low-coherence QPI approaches usually require spatially coherent illumination to ensure accurate and halo-free phase measurements \cite{17,18}, so the maximum achievable imaging resolution is still restricted to the coherent diffraction limit. Though synthetic aperture techniques via oblique \cite{19,20} or structured illumination \cite{21,22,23} can be used to extend the imaging resolution to that of their conventional, incoherent imaging counterparts, most of them require relatively complicated optical systems which are not typically available to most bio/pathologists, prohibiting their widespread use in biological and medical science.

On a different note, QPI can be realized through phase retrieval algorithms [specifically, iterative methods \cite{24,25,26,27,28,29,30} and transport of intensity equation (TIE) methods \cite{5,31,32,33,34}], which utilize only intensity measurements at multiple axially displaced planes (or under different illumination conditions) without explicit manipulation of object and reference beams. In contrast to iterative approaches, TIE is deterministic, requires less intensity measurements, and is fully compatible with commercially available microscopy units utilizing K\"ohler illumination optics \cite{5,34,35,36,37,38,39}. Besides, when the sample is illuminated with spatially partially coherent light from an optical condenser, TIE is able to achieve improved spatial resolution over the coherent diffraction limit as the angular content of the illumination contributes to the lateral resolution \cite{38}. Because of its numerical simplicity and flexible experimental configuration, TIE-based QPI approaches have found wide-ranging biomedical and technical applications, such as quantitative monitoring of cell growth in culture \cite{40,41}, investigations of cellular dynamics and toxin-mediated morphology changes \cite{33,34}, and characterization of optical elements \cite{42,43}. Although these works have demonstrated that it is possible to acquire accurate and speckle-free phase images under partially coherent illuminations, two fundamental limits still prevail: First, the phase reconstruction is strongly sensitive to low frequency artifacts, as a result of low frequency noise amplification in the direct inversion of the TIE (inverse Laplacian) \cite{44,45}. Second, the “washout" diffraction effect in partially coherent imaging system prevents the maximum possible resolution [diffraction limit of partially coherent imaging: the sum of the condenser and the objective numerical aperture (NA)] for phase imaging \cite{36,46}.

TIE provides a simple mathematical formalism to linearize the image formation process, linking the phase of an optical field to its intensity variations along the direction of propagation. Though it was originally derived under paraxial approximation and ideal imaging assumption, it is informative to reformulate the TIE phase reconstruction using the contrast transfer function (CTF) formalism \cite{38,44,45,46,47,48,49}, which is a primary tool to characterize the imaging system in both x-ray diffraction \cite{47,50} and optical microscopy communities \cite{51,52}. The major reason for its susceptibility to low-frequency noise lies in the fact that the phase transfer function of TIE, which relates the intensity measurement to the phase, goes to zero rapidly as the spatial frequency decreases in the small-defocus regime. On the other hand, in higher spatial frequency domain, the deviation between the TIE phase transfer function and the coherent CTF becomes significant, especially when the defocus distance is large, leading to attenuation and blurring in small-scale features\cite{45,50}. This is the so-called noise-resolution tradeoff in the choice of defocus distances for TIE phase retrieval. To alleviate such a problem, multi-distance approaches utilizing intensity measurements at both small and large defoci have been proposed \cite{45,53,54,55,56}, which allow the response of phase transfer function to be optimized over a wider range of spatial frequency at the expense of additional data acquisition and processing. These approaches have been further extended to account for partial coherence explicitly \cite{57,58,59}, by generalizing the coherent CTF to the weak object transfer function (WOTF) of a partially coherent system \cite{38,51,52}. Though it is well-known that there is no well-defined phase for partially coherent fields, the reconstructed ``phase'' from TIE has been proven to be connected to the transverse Poynting vector \cite{37} or Wigner distribution moment \cite{36}, and can be converted to the well-defined optical thickness of sample. Partial coherence extends the maximum achievable imaging resolution beyond coherent diffraction limit, but meanwhile, exacerbates the noise-to-resolution tradeoff because the decrease in illumination coherence tends to attenuate the phase effect \cite{36,57}. As is predicted by the WOTF analysis, the phase contrast progressively vanishes as the illumination NA approaches the objective NA, suggesting phase information can hardly be transferred into intensity via defocusing when illumination NA is large \cite{51,52,57,60}. For this reason, high quality, low-noise TIE-based QPI at resolution levels up to twice of coherent diffraction limit (diffraction limit of incoherent imaging) has not been reported so far.

In this work, we demonstrate the annular illumination based TIE (AI-TIE), which addresses the above-mentioned limitations in conventional TIE by replacing the conventional circular illumination aperture with an annular one. Although benefits of the annular illumination have been extensively studied in various fields of microscopy \cite{61,62,63,64} and microlithography \cite{65,66}, its potentials for resolution improvement and noise reduction in TIE imaging remains unexplored. By making use of the full NA range of imaging optics, a matched annular illumination aperture is able to significantly boost the phase contrast transfer function at low spatial frequencies, whilst improving the practical imaging resolution close to the incoherent diffraction limit. It is recognized that with matched annular illumination, the phase WOTF provides a broad Fourier coverage ($\sim$ two times of the objective NA) with a very flat response free from any zero-crossing in the small-defocus regime, which entirely removes the ill-posedness of the inversion and significantly suppresses the low-frequency noise and other reconstruction artifacts. Compared to conventional circular-aperture TIE phase imaging with multiple defocus distances, AI-TIE achieves improved imaging resolution (a gain close to 2 compared to coherent illumination, same as the one obtained using structured illumination or synthetic aperture holography \cite{67}), comparable noise-robustness, and higher imaging throughput due to less data acquisition (only three images required). The proposed method and theoretical conclusions are experimentally verified based on an off-the-shelf inverted microscope with appropriate annuli fitted into the condenser turret. The 208 nm transverse resolution achieved by combining high-NA annular illumination with high-NA objective detection reveals subcellular structures at high resolution in buccal epithelial cells. Time-lapse imaging of  \emph{in vitro} Hela cell samples is then presented, highlighting subcellular dynamics in mitosis and apoptosis in a non-invasive and label-free manner. The experimental results suggest that the developed AI-TIE is a promising, non-destructive, non-interventional tool for structural and functional cellular investigations. Because the approach can be implemented in widely accessible brightfield microscopy hardware, it is expected to be adopted on a large scale by non-specialists, such as bio/pathologists.

\section*{Materials and methods}

\subsection*{Image formation in a partially coherent microscope}

\begin{figure}[ht]
\centering
\includegraphics[width=0.95\linewidth]{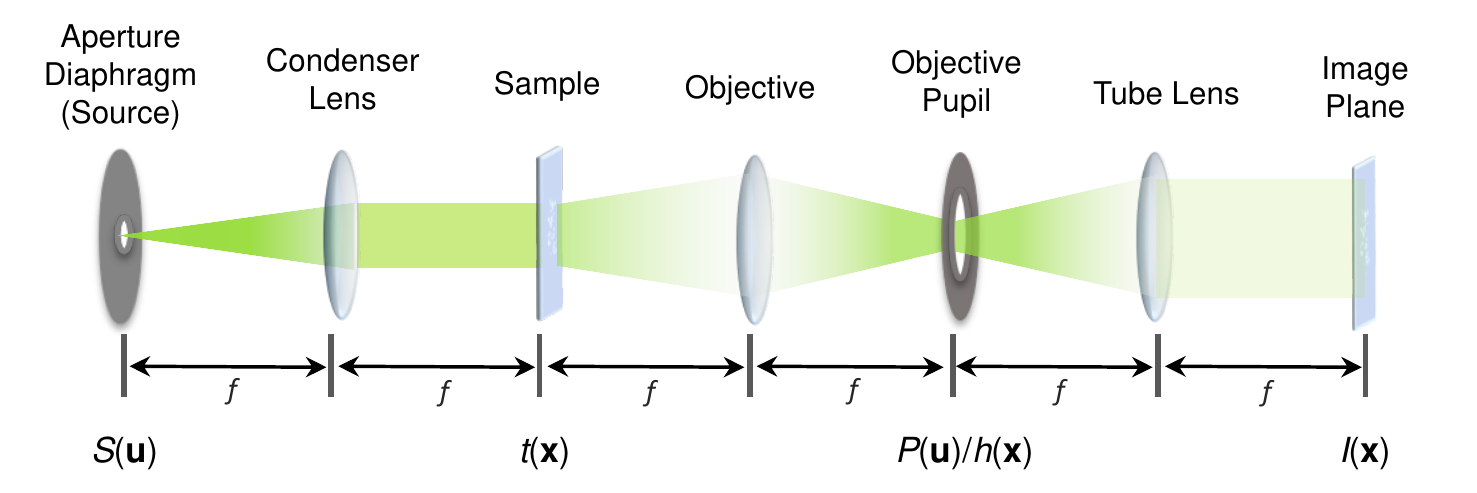}
\caption{Description of image formation in a partially coherent microscope.}
\label{fig1}
\end{figure}

\noindent
In general, the image formation in microscopic imaging systems can be described by Fourier transforms and a linear filtering operation in the pupil plane \cite{68}: a coherent imaging system is linear in complex amplitude, while an incoherent imaging system is linear in intensity. However, for partially coherent imaging, this process is complicated by the non-linear dependency of the image intensity on the object, light source, and imaging system. Considering the standard 6$f$ optical setup shown in Fig. \ref{fig1}, in which an ``incoherent'', delta-correlated light source (the condenser exit pupil) illuminates an object (with a complex transmittance $t\left( {\bf{x}} \right)$, where ${\bf{x}}$ is the two dimensional (2D) spatial coordinate in the real space) which is imaged using an objective lens. This configuration of the so-called K\"ohler illumination and telecentric detection can be applied to almost all brightfield microscopic configurations \cite{39,51}. In the following, the formal description of the imaging is further limited to quasi-monochromatic light and unit (1.0$\times$) magnification. An extension to polychromatic illuminations and arbitrary magnification is straightforward.

Considering the illumination emerging from a light source with intensity distribution $S\left( {\bf{u}} \right)$ in the aperture diaphragm plane (where ${\bf{u}}$ is the corresponding 2D coordinate in the Fourier space), the image of the object captured at the image plane can be described by (see \textbf{Supplementary Information A} for detailed derivation):
\begin{equation}\label{1}
I\left( {\bf{x}} \right) = \int {S\left( {\bf{u}} \right)} {\left| {\int {t\left( {{\bf{x'}}} \right)h\left( {{\bf{x}} - {\bf{x'}}} \right){e^{i2\pi {\bf{ux'}}}}d{\bf{x'}}} } \right|^2}d{\bf{u}} \equiv \int {S\left( {\bf{u}} \right)} {I_{\bf{u}}}\left( {\bf{x}} \right)d{\bf{u}}
\end{equation}
where $h$  represents the amplitude point spread function (PSF) of the imaging system. Equation (\ref{1}) suggests that image formation for the optical imaging with K\"ohler illumination can be interpreted as an incoherent superposition of all intensities of the coherent partial images ${I_{\bf{u}}}\left( {\bf{x}} \right)$ arising from all light source points:
\begin{equation}\label{2}
{I_{\bf{u}}}\left( {\bf{x}} \right) = {\left| {\int {t\left( {{\bf{x'}}} \right)h\left( {{\bf{x}} - {\bf{x'}}} \right){e^{i2\pi {\bf{ux'}}}}d{\bf{x'}}} } \right|^2}
\end{equation}
Equation (\ref{1}) can also be rewritten in terms of their respective Fourier transforms:
\begin{equation}\label{3}
I\left( {\bf{x}} \right) =  \iiint{  S\left( {\bf{u}} \right)T\left( {{{\bf{u}}_1}} \right){T^{\rm{*}}}\left( {{{\bf{u}}_2}} \right)P\left( {{\bf{u}}{\rm{ + }}{{\bf{u}}_1}} \right){P^{\rm{*}}}\left( {{\bf{u}}{\rm{ + }}{{\bf{u}}_2}} \right){e^{i2\pi {\bf{x}}\left( {{{\bf{u}}_1}{\rm{ - }}{{\bf{u}}_2}} \right)}}d{{\bf{u}}_1}d{{\bf{u}}_2}d{\bf{u}}}
\end{equation}
where $P\left( {\bf{u}} \right){\rm{ = }}\left| {P\left( {\bf{u}} \right)} \right|{{\mathop{\rm e}\nolimits} ^{ikW\left( {\bf{u}} \right)}}$  is the coherent transfer function with the pupil function  $\left| {P\left( {\bf{u}} \right)} \right|$ (in the most case is a circ-function) and the wave-front aberration $W\left( {\bf{u}} \right)$ . Due to superposition of intensities, the observed intensity is not linear in the specimen transmittance. The intensity spectrum consists of mixing of pairs of spatial frequencies in the amplitude spectrum of the specimen. Such a dependence of the image intensity on amplitude of the specimen is called `bi-linear' (i.e., linear in pairs) dependence. Separating the contribution of the specimen and system leads to the notion of the transmission cross-coefficient (TCC) \cite{69,70}
\begin{equation}\label{4}
TCC\left( {{{\bf{u}}_1},{{\bf{u}}_2}} \right) = \iint{ {S\left( {\bf{u}} \right)P\left( {{\bf{u}}{\rm{ + }}{{\bf{u}}_1}} \right){P^{\rm{*}}}\left( {{\bf{u}}{\rm{ + }}{{\bf{u}}_2}} \right)d{\bf{u}}}}
\end{equation}
We now obtain the transfer function for partially coherent image formation:
\begin{equation}\label{5}
I\left( {\bf{x}} \right) =   \iint{ {T\left( {{{\bf{u}}_1}} \right){T^{\rm{*}}}\left( {{{\bf{u}}_2}} \right)TCC\left( {{{\bf{u}}_1},{{\bf{u}}_2}} \right){e^{i2\pi {\bf{x}}\left( {{{\bf{u}}_1}{\rm{ - }}{{\bf{u}}_2}} \right)}}d{{\bf{u}}_1}d{{\bf{u}}_2}}}
\end{equation}
The image intensity is therefore computed by the linear superposition of the interference patterns of plane waves with spatial frequencies  ${{\bf{u}}_1}$ and   ${{\bf{u}}_2}({e^{i2\pi {\bf{x}}\left( {{{\bf{u}}_1}{\rm{ - }}{{\bf{u}}_2}} \right)}})$, and amplitude given by the TCC and the spatial object spectrum contribution at ${{\bf{u}}_1}$  and  ${{\bf{u}}_2}$ ($T\left( {{{\bf{u}}_1}} \right){T^{\rm{*}}}\left( {{{\bf{u}}_2}} \right)$).

\subsection*{WOTF of a partially coherent microscope}

\begin{figure}[ht]
\centering
\includegraphics[width=0.95\linewidth]{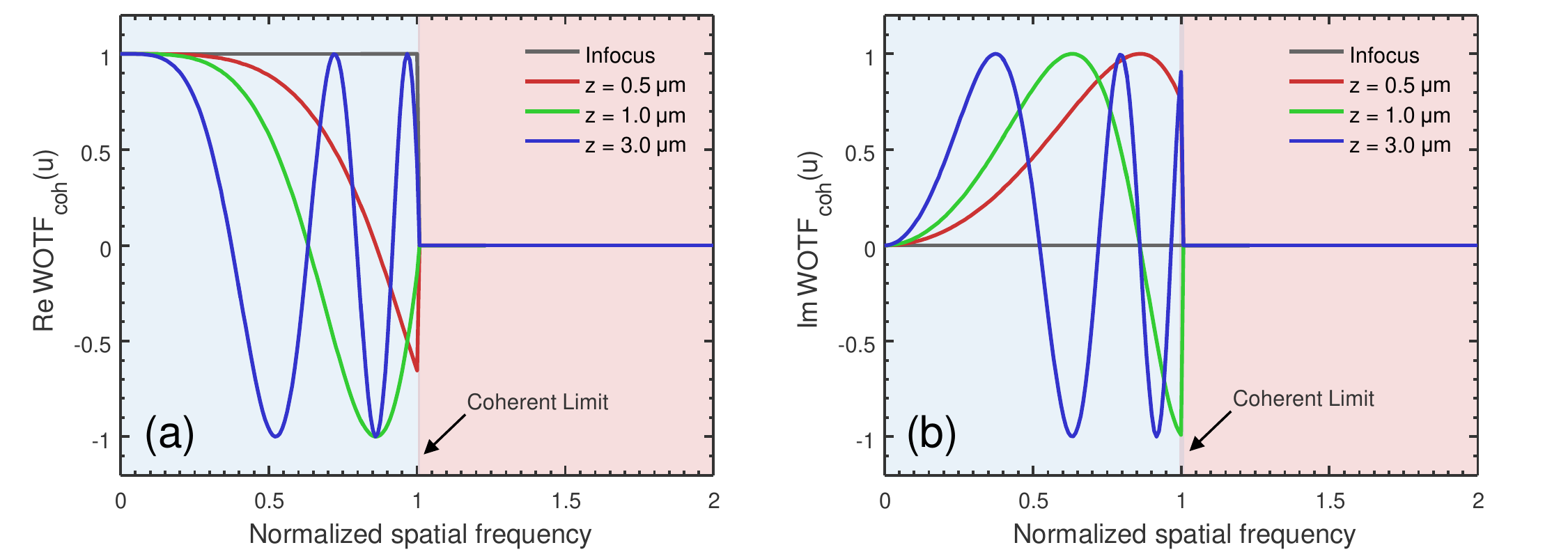}
\caption{The $WOT{F_{{\rm{coh}}}}$ for various defocus distances ($N{A_{obj}}{\rm{ =  }}0.8$, $\lambda  = 550$nm, the spatial frequency coordinate is normalized against the coherent resolution limit ${N{A_{obj}}\slash\lambda}$). (a) Real part of $WOT{F_{{\rm{coh}}}}$ (amplitude CTF); (b) Imaginary part of $WOT{F_{{\rm{coh}}}}$ (phase CTF).}
\label{fig2}
\end{figure}

\noindent
The image formation in partially coherent systems is not linear in either amplitude or intensity but is bilinear, which makes image restoration and phase recovery complicated. To simplify the mathematical formulation, the weak object approximation is often applied to linearize the phase retrieval problem \cite{71,72}. The complex transmittance of a weak object can be represented as
\begin{equation}\label{6}
t\left( {\bf{x}} \right) \equiv a\left( {\bf{x}} \right){{\mathop{\rm e}\nolimits} ^{i\phi \left( {\bf{x}} \right)}} \approx a\left( {\bf{x}} \right)\left[ {1 + i\phi \left( {\bf{x}} \right)} \right]\mathop  \approx \limits^{a\left( {\bf{x}} \right) = {a_0}{\rm{ + }}\Delta a\left( {\bf{x}} \right)} {a_0} + \Delta a\left( {\bf{x}} \right) + i{a_0}\phi \left( {\bf{x}} \right)
\end{equation}
where $a\left( {\bf{x}} \right)$  is the absorption distribution with a mean value of ${a_0}$; $\phi \left( {\bf{x}} \right)$ is the phase distribution. In Eq. (\ref{5}), the interference terms of the object function with spatial frequencies ${{\bf{u}}_1}$  and ${{\bf{u}}_2}$  can be approximated as:
\begin{equation}\label{7}
T\left( {{{\bf{u}}_1}} \right){T^{\rm{*}}}\left( {{{\bf{u}}_2}} \right) = a_0^2\delta \left( {{{\bf{u}}_1}} \right)\delta \left( {{{\bf{u}}_2}} \right) + {a_0}\delta \left( {{{\bf{u}}_2}} \right)\left[ {\Delta \tilde a\left( {{{\bf{u}}_1}} \right) + i{a_0}\tilde \phi \left( {{{\bf{u}}_1}} \right)} \right] + {a_0}\delta \left( {{{\bf{u}}_1}} \right)\left[ {\Delta {{\tilde a}^{\rm{*}}}\left( {{{\bf{u}}_2}} \right) - i{a_0}{{\tilde \phi }^{\rm{*}}}\left( {{{\bf{u}}_2}} \right)} \right]
\end{equation}
Note that the interference of scattered light with scattered light (weak) is neglected, which just corresponds to the first order Born approximation commonly used in diffraction tomography \cite{51,73,74}. Substituting Eq. (\ref{7}) to Eq. (\ref{5}) yields the intensity of the partially coherent image for a weak object:
\begin{equation}\label{8}
I\left( {\bf{x}} \right) = a_0^2TCC\left( {{\bf{0}},{\bf{0}}} \right) + 2{a_0}{\mathop{\rm Re}\nolimits} \left\{ {\int {TCC\left( {{\bf{u}},{\bf{0}}} \right)\left[ {\Delta \tilde a\left( {\bf{u}} \right){\rm{ + i}}{a_0}\tilde \phi \left( {\bf{u}} \right)} \right]{e^{i2\pi {\bf{xu}}}}d{\bf{u}}} } \right\}
\end{equation}
where the simple relation $TC{C^{\rm{*}}}\left( {{\bf{0}},{\bf{u}}} \right){\rm{ = }}TCC\left( {{\bf{u}},{\bf{0}}} \right)$  is used (TCC is Hermitian symmetric). It now becomes obvious that the image contrast due to the absorption and phase are decoupled and linearized. In the following, we denote the $TCC\left( {{\bf{u}},{\bf{0}}} \right)$, the linear part of the TCC, as weak object transfer function (WOTF) \cite{49,52}:
\begin{equation}\label{9}
WOTF\left( {\bf{u}} \right) \equiv TCC\left( {{\bf{u}},0} \right) =  \iint{{S\left( {{\bf{u'}}} \right)P\left( {{\bf{u'}}{\rm{ + }}{\bf{u}}} \right){P^{\rm{*}}}\left( {{\bf{u'}}} \right)d{\bf{u'}}}}
\end{equation}
For an aberration-free system with axisymmetric illumination and objective pupil, the WOTF is real and even, which gives only absorption contrast but no phase contrast, regardless of the degree of illumination coherence. Breaking the symmetry in either source or pupil function results in phase contrast, just as differential phase-contrast microscopy \cite{75,76,77}, pyramid wavefront sensor \cite{78,79}, and partitioned/programmable aperture microscopy \cite{80,81}. Another more convenient way of producing phase contrast is to introduce an imaginary part into the transfer function by defocusing the optical system:
\begin{equation}\label{10}
P\left( {\bf{u}} \right){\rm{ = }}\left| {P\left( {\bf{u}} \right)} \right|{e^{jk{\rm{z}}\sqrt {1 - {\lambda ^2}{{\left| {\bf{u}} \right|}^2}} }}, \quad \left| {\bf{u}} \right| \le {\lambda ^{ - 1}}
\end{equation}
where $z$  is the defocus distance along the optical axis. Substituting the complex pupil function into Eq. (\ref{9}) results in a complex (but even) WOTF:
\begin{equation}\label{11}
WOTF\left( {\bf{u}} \right) =   \iint{ {s\left( {{\bf{u'}}} \right)\left| {P\left( {{\bf{u'}}} \right)} \right|\left| {P\left( {{\bf{u'}} + {\bf{u}}} \right)} \right|{e^{jk{\rm{z}}\left( { - \sqrt {1 - {\lambda ^2}{{\left| {{\bf{u'}}} \right|}^2}} {\rm{ + }}\sqrt {1 - {\lambda ^2}{{\left| {{\bf{u}} + {\bf{u'}}} \right|}^2}} } \right)}}d{\bf{u'}}}}
\end{equation}
The transfer functions for absorption and phase component of a weak object is then given by the real and imagery part of WOTF, respectively:
\begin{equation}\label{12}
{H_A}\left( {\bf{u}} \right) = 2{a_0}{\mathop{\rm Re}\nolimits} \left[ {WOTF\left( {\bf{u}} \right)} \right]
\end{equation}
\begin{equation}\label{13}
{H_P}\left( {\bf{u}} \right) = 2a_0^2{\mathop{\rm Im}\nolimits} \left[ {WOTF\left( {\bf{u}} \right)} \right]
\end{equation}
In the coherent limit ($S\left( {\bf{u}} \right){\rm{ = }}\delta \left( {\bf{u}} \right)$), the WOTF can be greatly simplified:
\begin{equation}\label{14}
WOT{F_{{\rm{coh}}}}\left( {\bf{u}} \right) = {\left| {P\left( {\bf{u}} \right)} \right|^2}{e^{jk{\rm{z}}\left( {\sqrt {1 - {\lambda ^2}{{\left| {\bf{u}} \right|}^2}}  - 1} \right)}}
\end{equation}
where the exponential term just corresponds to the propagation kernel for the angular spectrum representation of a coherent complex field. Note that the real part (for absorption contrast) and imaginary part (for phase contrast) of the WOTF are given by the cosine and sine of $jk{\rm{z}}\left( {\sqrt {1 - {\lambda ^2}{{\left| {\bf{u}} \right|}^2}}  - 1} \right)$, respectively. By further invoking the paraxial approximation ($jk{\rm{z}}\left( {\sqrt {1 - {\lambda ^2}{{\left| {\bf{u}} \right|}^2}}  - 1} \right) \approx  - i\pi \lambda {\rm{z}}{\left| {\bf{u}} \right|^2}$), the WOTF can be written as
\begin{equation}\label{15}
WOT{F_{{\rm{coh}}}}\left( {\bf{u}} \right) = {\left| {P\left( {\bf{u}} \right)} \right|^2}\left[ {\cos \left( {\pi \lambda z{{\left| {\bf{u}} \right|}^2}} \right) - i\sin \left( {\pi \lambda z{{\left| {\bf{u}} \right|}^2}} \right)} \right]
\end{equation}
The real and imaginary parts of $WOT{F_{{\rm{coh}}}}$ are shown in Fig. \ref{fig2} for various defocus distances. It should be noted that apart from the aperture function ${\left| {P\left( {\bf{u}} \right)} \right|^2}$ , the imaginary part of the $WOT{F_{{\rm{coh}}}}$  (the sine term) corresponds to the phase CTF that is widely applied in the field of propagation-based coherent X-ray diffraction imaging \cite{47,50,71}. For weak defocus, the sine term can be further approximated by a parabolic function $\sin \left( {\pi \lambda z{{\left| {\bf{u}} \right|}^2}} \right) \approx \pi \lambda z{\left| {\bf{u}} \right|^2}$,  which is a Laplacian in Fourier space, corresponding to the phase transfer function implied by TIE in the case of smooth intensity ($ - k\frac{{\partial I}}{{\partial z}} = I{\nabla ^2}\phi $) \cite{33,50,55,58}.

\begin{figure}[htb!]
\centering
\includegraphics[width=0.85\linewidth]{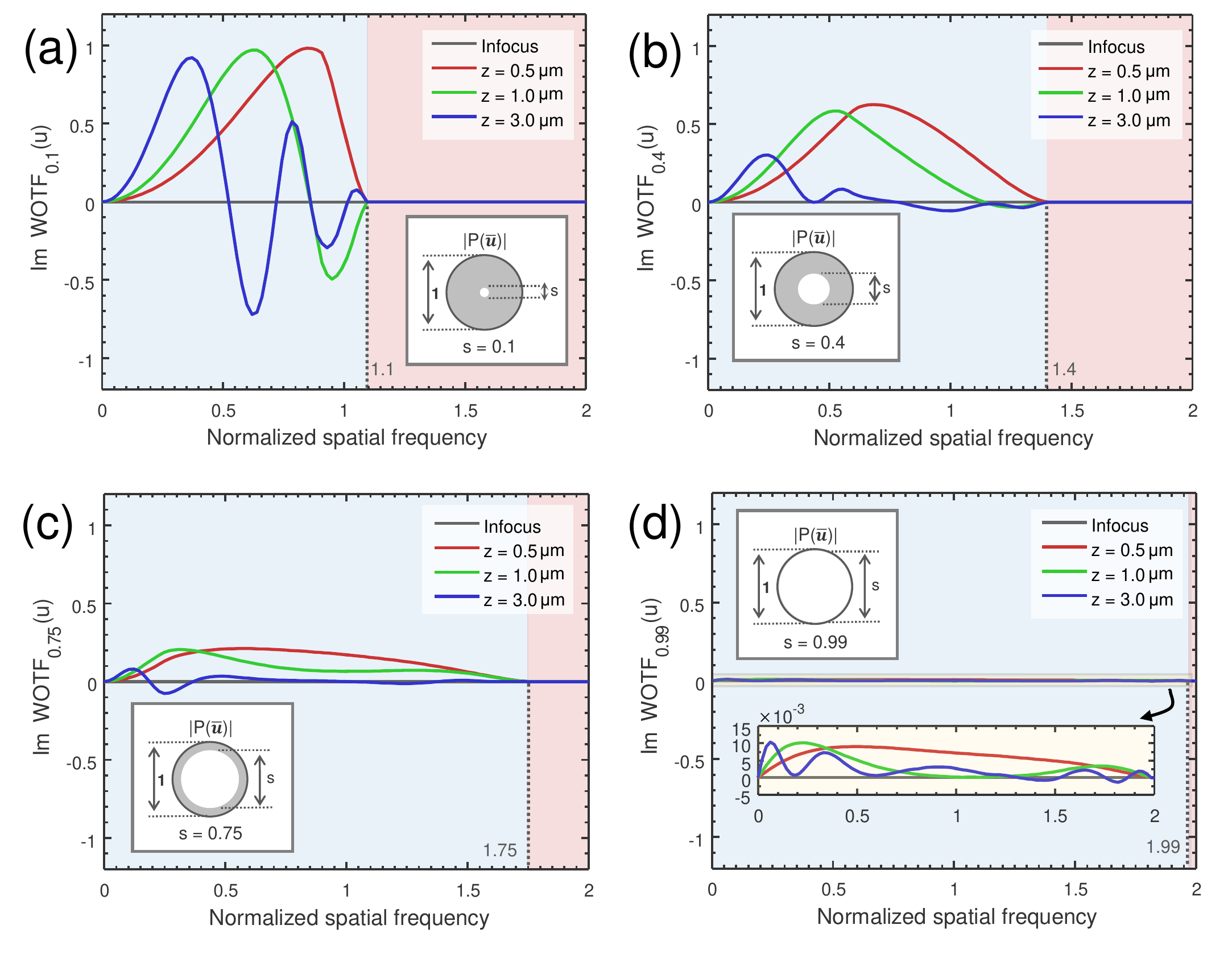}
\caption{The imaginary part of WOTF (phase WOTF) for various coherent parameters and defocus distances ( $N{A_{obj}}{\rm{ =  }}0.8$, $\lambda  = 550$nm, the spatial frequency coordinate is normalized against the coherent resolution limit ${N{A_{obj}}\slash\lambda}$). (a) \emph{s} = 0.1; (b) \emph{s} = 0.4; (c) \emph{s} = 0.75; (d) \emph{s} = 0.99.}
\label{fig3}
\end{figure}

On the other hand, if the light source is larger than the pupil diameter [i.e. $s \ge 1$, where \emph{s} is the so-called coherence parameter \cite{38,49,51,52}, which is the ratio of illumination to objective NA, $s = {{N{A_{ill}}} \mathord{\left/
 {\vphantom {{N{A_{ill}}} {N{A_{obj}}}}} \right. \kern-\nulldelimiterspace} {N{A_{obj}}}}$ ], the system behaves the same as an incoherent microscope [but note that it is not really incoherent (which requires $s \to \infty $ ) because a weak object is assumed in this case] with the $WOT{F_{{\rm{incoh}}}}$  given by the autocorrelation of the pupil function:
\begin{equation}\label{16}
WOT{F_{{\rm{incoh}}}}\left( {\bf{u}} \right) =  \iint{{P\left( {{\bf{u'}}{\rm{ + }}{\bf{u}}} \right){P^{\rm{*}}}\left( {{\bf{u'}}} \right)d{\bf{u'}}}}
\end{equation}
Although the spatial frequency cut-off of an incoherent system is twice that of a coherent system, the $WOT{F_{{\rm{incoh}}}}$  is always real, and hence the phase part of the specimen cannot be imaged (see Supplementary Figure S1). For this reason, partially incoherent illumination with a reduced light source dimension ($s < 1$ ) is generally used to provide phase contrast under a normal brightfield microscope platform. Figure \ref{fig3} shows the imaginary part of WOTF (phase WOTF) for various coherent parameters and defocus distances (see also Supplementary Figure S2 for the real part of WOTF). These results are numerically calculated from Eq. (\ref{11}) directly, as a weighted area of overlap over $S\left( {{\bf{u'}}} \right){P^{\rm{*}}}\left( {{\bf{u'}}} \right)$  and $P\left( {{\bf{u'}}{\rm{ + }}{\bf{u}}} \right)$  (when the system is in-focus, the WOTF can be represented analytically, as given in the \textbf{Supplementary Information B}). Note that all WOTF curves shown here and in the following are normalized by the background intensity $WOTF\left( {\bf{0}} \right)$ (intensity of the source integrated over the pupil), thus demonstrating the ``contrast'' of the image (relative strengths of information-bearing portion of the image and the ever-present background). It can be seen from Figs. \ref{3} and S2 that though the spatial frequency limit of the WOTF can be extended to $1 + s$  compared with the coherent case, increasing coherent parameter in general diminishes the response of phase contrast (imaginary part) considerably. This is in consistent with the well-known phenomenon that the condenser of a microscope must be stopped down to produce an appreciable contrast for phase information. But in this case, the associated reduction in the spatial frequency cut-off becomes a major issue, leading to a sub-optimal imaging resolution. Thus, as suggested in many literatures, for defocus-based QPI like the TIE, a compromise on illumination coherence has to be made to balance the imaging resolution and the phase contrast versus defocus \cite{34,36,58}. It should be also noted that though the defocused WOTF for partially coherent imaging was previously derived under paraxial and weak defocus assumptions \cite{52}, it is still valid for non-paraxial and large defocus conditions despite that no analytic expressions can be obtained.

\subsection*{Quantitative phase reconstruction by inversion of TIE and WOTF }

\noindent
TIE provides a simple and general method for QPI by only measuring the intensity along the beam propagation direction \cite{31}:
\begin{equation}\label{17}
 - k\frac{{\partial I\left( {\bf{x}} \right)}}{{\partial z}} = \nabla  \cdot \left[ {I\left( {\bf{x}} \right)\nabla \phi \left( {\bf{x}} \right)} \right]
\end{equation}
where $\nabla$ is the transverse gradient operator and $\mathbf{\cdot}$ denotes the dot product. The ${{\partial I} \mathord{\left/{\vphantom {{\partial I} {\partial z}}} \right.\kern-\nulldelimiterspace} {\partial z}}$  is the axial derivative of the intensity of the field, which can be approximated by finite differences. Normally, the solution of the TIE involves an auxiliary function $\nabla \psi \left( {\bf{x}} \right) = I\left( {\bf{x}} \right)\nabla \phi \left( {\bf{x}} \right)$  to convert it into the following two Poisson equations \cite{25,31,37,82}:
\begin{equation}\label{18}
- k\frac{{\partial I\left( {\bf{x}} \right)}}{{\partial z}} = {\nabla ^2}\psi \left( {\bf{x}} \right)
\end{equation}
\begin{equation}\label{19}
\nabla \psi \left( {\bf{x}} \right) = I\left( {\bf{x}} \right)\nabla \phi \left( {\bf{x}} \right)
\end{equation}
By solving the first Poisson equation [Eq. (\ref{18})], we can get the solution to the auxiliary function $\psi \left( {\bf{x}} \right)$ , thus the phase gradient can be obtained ( $\nabla \phi \left( {\bf{x}} \right) = {I^{ - 1}}\left( {\bf{x}} \right)\nabla \psi \left( {\bf{x}} \right)$). Solving the second Poisson equation [Eq. (\ref{19})] allows for phase integration. When the intensity variation along the transversal dimension is small (which is generally true for unstained cells and tissues), TIE can be simplified as only one Poisson equation
\begin{equation}\label{20}
 - k\frac{{\partial I\left( {\bf{x}} \right)}}{{\partial z}} = I\left( {\bf{x}} \right){\nabla ^2}\phi \left( {\bf{x}} \right)
\end{equation}
and the two-step solution simply boils down to an inverse Laplacian in Fourier space ($1/\pi \lambda z{\left| {\bf{u}} \right|^2}$), which is equivalent to an inversion of weak-defocus CTF or WOTF in the coherent limit \cite{33,50,55,58}.

One limitation of TIE is that it only gives the phase of the `image' (not the object) since no parameters about the imaging system is involved in this equation \cite{39,52}. To explicitly account for the effect of partial coherence and imaging system, the Fourier-space Laplacian ($\pi \lambda z{\left| {\bf{u}} \right|^2}$) should be replaced by the WOTF. It is straightforward to deduce that once the objective and source function is circularly symmetric, the WOTF [Eq. (\ref{11})] is a Hermitian function of defocus: the real part of WOTF (for amplitude contrast) is always an even function while the imaginary part (for phase contrast) is always an odd function with respect to $z$ . Thus, similar to the longitudinal intensity derivative in TIE phase retrieval, for the inversion of WOTF, one needs to capture two intensity images with equal and opposite defoci $ \pm \Delta z$. Then addition of the two images gives a pure amplitude-contrast image, while subtraction of the two images gives a pure phase-contrast image \cite{38,52}. Besides, the in-focus image ${I_0}\left( {\bf{x}} \right)$  is taken as the background intensity, $a_0^2TCC\left( {{\bf{0}},{\bf{0}}} \right)$ , which is used for normalizing the difference of the two defocused images
\begin{equation}\label{21}
\frac{{{{\tilde I}_{\Delta {\rm{z}}}}\left( {\bf{u}} \right) - {{\tilde I}_{ - \Delta {\rm{z}}}}\left( {\bf{u}} \right)}}{{4{{\tilde I}_0}\left( {\bf{u}} \right)}} = {\mathop{\rm Im}\nolimits} \left[ {WOTF\left( {\bf{u}} \right)} \right]\tilde \phi \left( {\bf{u}} \right)
\end{equation}
leading to a linear relation between the phase and the WOTF. Quantitative phase information can then be reconstruction by Fourier space de-convolution in one-step \cite{33,48}.
\begin{equation}\label{22}
\phi \left( {\bf{x}} \right) = {\mathscr{F}^{ - 1}}\left\{ {\frac{{{{\tilde I}_{\Delta {\rm{z}}}}\left( {\bf{u}} \right) - {{\tilde I}_{ - \Delta {\rm{z}}}}\left( {\bf{u}} \right)}}{{4{{\tilde I}_0}\left( {\bf{u}} \right)}}\frac{{{\mathop{\rm Im}\nolimits} \left[ {WOTF\left( {\bf{u}} \right)} \right]}}{{{{\left| {{\mathop{\rm Im}\nolimits} \left[ {WOTF\left( {\bf{u}} \right)} \right]} \right|}^2}{\rm{ + }}\alpha }}} \right\}
\end{equation}
where  ${\mathscr{F}^{ - 1}}$ represents the inverse Fourier transform and $\alpha$  is the Tikhonov-regularization parameter which is used to avoid the numerical instability of the `division by zero’ (over-amplification of noise) that would arise at frequencies where the response of  ${\mathop{\rm Im}\nolimits} \left[ {WOTF\left( {\bf{u}} \right)} \right]$ is close to zero \cite{45,83}. The advantage of introducing WOTF theory into TIE is that it can derive the phase of the object rather than that of the image by considering the effect of illumination coherence and imaging system explicitly \cite{33,39,48,52,58}. For coherent imaging, the WOTF becomes the coherent CTF limited by a pupil function [Eq. (\ref{14})]. While for partially coherent imaging, the WOTF should be calculated with the knowledge about the coherent parameter of the imaging system. Different from the conventional TIE method, the WOTF is still valid beyond the paraxial and weak-defocus regime if a rigorous treatment of defocusing based on angular spectrum propagation (instead of Fresnel propagation) is employed [as in Eqs.(\ref{10}) and (\ref{11})].

Partially coherent illumination is particularly useful in QPI, providing promising advantages of speckle-free imaging, greater light throughput, and enhanced imaging resolution (1+s times improvement in resolution limit over coherent imaging). However, the actual performance of the phase reconstruction is heavily dependent on the form of WOTF. As shown in Fig. (\ref{fig3}), the phase WOTF tends to have very low response at both low and high spatial frequencies, which compromises the signal-to-noise ratio (SNR) and the maximum achievable resolution of the phase reconstruction. Thus, the low-frequency noise and the high-frequency blurring are two major issues for TIE and other related defocus-based QPI approaches. These problems can be partially ameliorated by using multi-distance ($\geq$2) approaches in which the response of phase transfer function can be optimized over a wider range of spatial frequencies by proper frequency selection/combination \cite{45,55,56,57}, or least-squares weighting \cite{50,58}. However, the increase in the number of measurements prolongs the data acquisition and processing times, limiting the throughput of the imaging system. Moreover, these multi-distance approaches are impotent to improve the imaging resolution when \emph{s} is large, in which case the phase contrast becomes too weak to be usable. As is shown in Figs. \ref{fig3} and S2, when the illumination NA approaches the objective NA, the phase contrast gradually disappears, suggesting that the phase information is lost in the captured intensity image. This is the major reason why high quality, low-noise TIE-based QPI with imaging resolution up to the incoherent diffraction limit has not been reported yet.

\subsection*{WOTF optimization with high-NA annular illumination}

\noindent
In the previous research and the above derivations, the illumination aperture $S\left( {\bf{u}} \right)$ is only restricted to circular shapes. This assumption is well coincident with the K\"ohler illumination configuration, in which a variable circular diaphragm is used for controlling the spatial coherence (coherence parameter \emph{s}) of the illumination. However, it should be also noted that the shape of condenser aperture diaphragm can also be manipulated to optimize the form of WOTF, thus allowing for improved phase imaging performance. Let us consider that the source becomes annular in shape and the objective pupil is still a normal circle:

\begin{equation}\label{23}
S\left( {{\bf{\bar u}}} \right) = \left\{ \begin{array}{l}
1,{\kern 1pt} {\kern 1pt} {\kern 1pt} {\kern 1pt} {\kern 1pt} {\kern 1pt} {\kern 1pt} {\kern 1pt} {\kern 1pt} {\kern 1pt} {\kern 1pt} {\kern 1pt} {\kern 1pt} {\kern 1pt} {\kern 1pt} {\kern 1pt} {\kern 1pt} {\kern 1pt} {\kern 1pt} {s_1} \le {\kern 1pt} \left| {{\bf{\bar u}}} \right| \le {s_2}\\
0,{\kern 1pt} {\kern 1pt} {\kern 1pt} {\kern 1pt} {\kern 1pt} {\kern 1pt} {\kern 1pt} {\kern 1pt} {\kern 1pt} {\kern 1pt} \left| {{\bf{\bar u}}} \right| < {s_1},{\kern 1pt} {\kern 1pt} {\kern 1pt} \left| {{\bf{\bar u}}} \right| > {s_2}
\end{array} \right.
\end{equation}
\begin{equation}\label{24}
\left| {P\left( {{\bf{\bar u}}} \right)} \right| = \left\{ \begin{array}{l}
1,{\kern 1pt} {\kern 1pt} {\kern 1pt} {\kern 1pt} {\kern 1pt} {\kern 1pt} {\kern 1pt} {\kern 1pt} {\kern 1pt} {\kern 1pt} {\kern 1pt} {\kern 1pt} {\kern 1pt} {\kern 1pt} {\kern 1pt} {\kern 1pt} {\kern 1pt} {\kern 1pt} \left| {{\bf{\bar u}}} \right| \le 1\\
0,{\kern 1pt} {\kern 1pt} {\kern 1pt} {\kern 1pt} {\kern 1pt} {\kern 1pt} {\kern 1pt} {\kern 1pt} {\kern 1pt} {\kern 1pt} {\kern 1pt} {\kern 1pt} {\kern 1pt} {\kern 1pt} {\kern 1pt} {\kern 1pt} {\kern 1pt} \left| {{\bf{\bar u}}} \right| > 1
\end{array} \right.
\end{equation}
\begin{figure}[!b]
\centering
\includegraphics[width=0.85\linewidth]{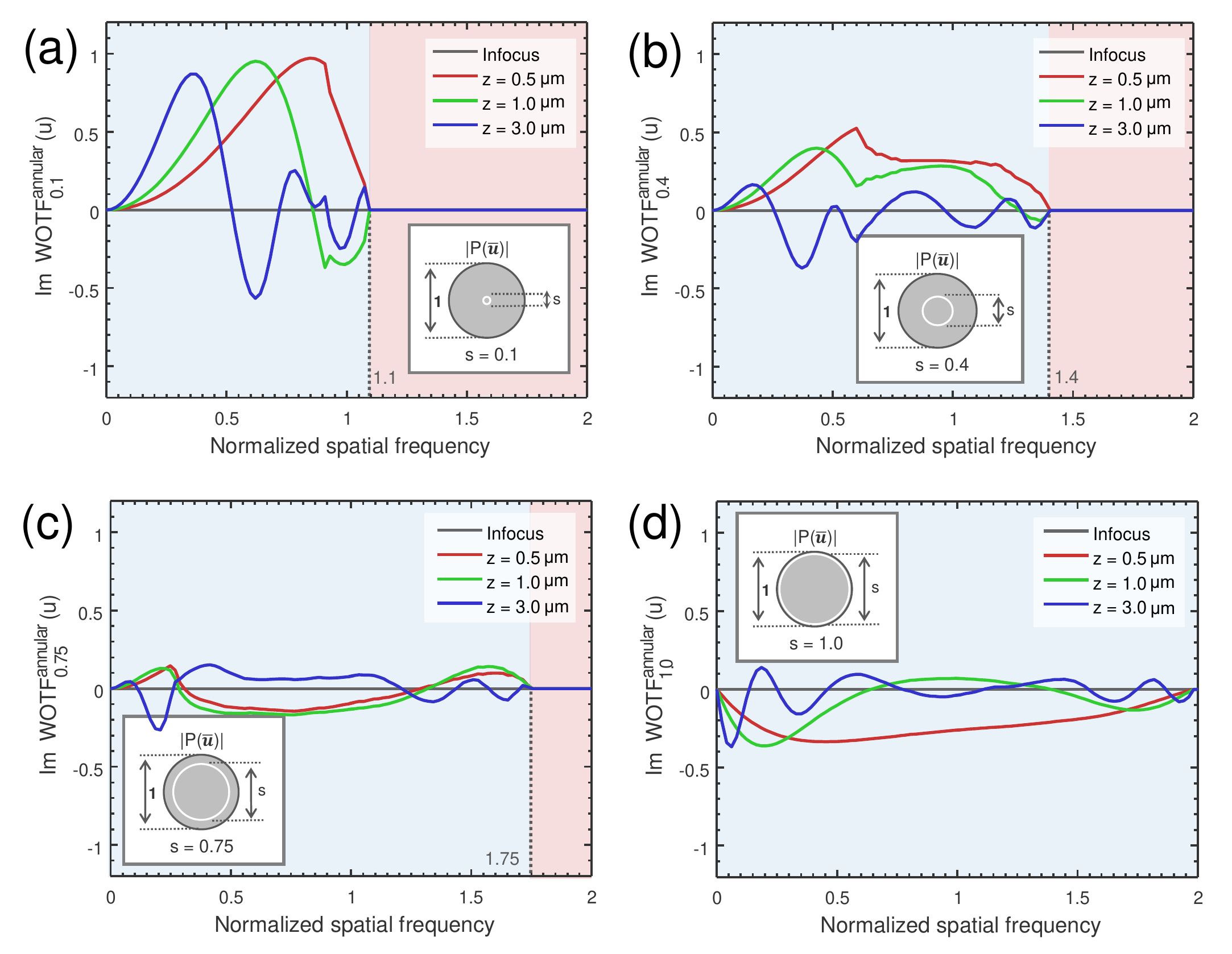}
\caption{The imaginary part of WOTF (phase WOTF) when the source is a narrow annulus ($N{A_{obj}}{\rm{ =  }}0.8$, $\lambda  = 550$nm, the spatial frequency coordinate is normalized against the coherent resolution limit ${N{A_{obj}}\slash\lambda}$). (a) \emph{s} = 0.1; (b) \emph{s} = 0.4; (c) \emph{s} = 0.75; (d) \emph{s} = 1.0.}
\label{fig4}
\end{figure}
It should be noted in Eq. (\ref{23}) and (\ref{24}) that the frequency coordinate ${\bf{\bar u}}$  is normalized to the coherent diffraction limit ($N{A_{obj}}\slash\lambda$), and the ${s_1}$  and ${s_2}$ represent the NAs of the inner and outer circles of the annulus, respectively. In order to achieve high resolution, the cut-off spatial frequency must be as high as possible, and to achieve good SNR, the phase WOTF should have a large magnitude over a wide range of spatial frequencies. In the following of this section, we will examine the form of WOTF under annular illumination and determine the optimum aperture function for high-resolution QPI.

\begin{figure}[htb!]
\centering
\includegraphics[width=0.85\linewidth]{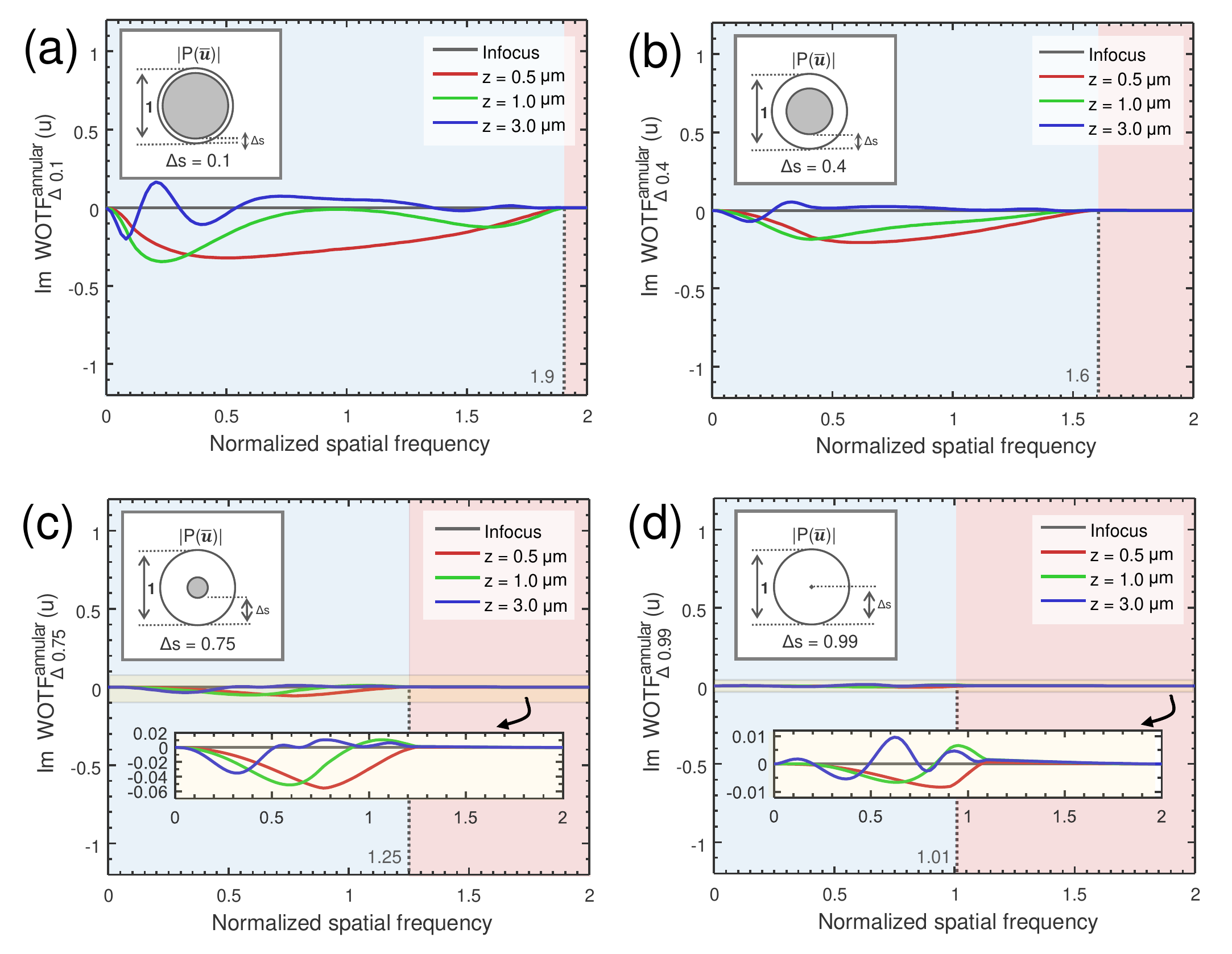}
\caption{The imaginary part of WOTF (phase WOTF) when the thickness of the annulus varies ($N{A_{obj}}{\rm{ =  }}0.8$, $\lambda  = 550nm$ , the spatial frequency coordinate is normalized against the coherent resolution limit ${N{A_{obj}}\slash\lambda}$). (a) $\Delta s$ = 0.1; (b) $\Delta s$ = 0.4; (c) $\Delta s$ = 0.75; (d) $\Delta s$  = 0.99.}
\label{fig5}
\end{figure}

We first consider the case when $s = {s_1} \approx {s_2}$ , which makes the source a narrow annulus. The WOTFs are plotted for various values of $s$ and defocus distances in Fig. \ref{fig4} (the in-focus WOTF can be represented analytically, as given in the \textbf{Supplementary Information B}). When these curves are compared to those shown in Fig. \ref{fig3}, it may immediately be observed that for $s = 0.1$, the response tends to quite similar with the case of circular illumination, especially for low spatial frequencies. However, with the increase in $s$ , the imaging properties of annular illumination begin to differ from those of circular illuminations: the phase contrast no longer vanishes when   approaches to one. Most notably, when  $s = 1$ and the defocus distance is small (0.5 $\mu$m), the WOTF demonstrates strong but negative (inverse) phase contrast. The phase WOTF contains no zero-crossings (except the origin), and tends to be roughly constant across a wide range of frequencies ( 0 to ${N{A_{obj}}\slash\lambda}$ ), which is an ideal form for noise-robust, high-resolution, and well-posed phase reconstruction. However, the practical annulus cannot be infinitesimally thin. In Fig. \ref{fig5}, we further demonstrate the effect of varying the thickness of the annulus by fixing the NA of the outer circles to be 1 (${s_2} = 1$) and only changing the thickness of the annulus ( ${s_1} = \Delta s$ from 0 to 1). As might be expected, the phase contrast is reduced as the annulus width increases. When $\Delta \emph{s} \to 1$ , the phase contrast finally goes to zero, which is just the incoherent case of the circular illumination [Eq. (\ref{16}) is reproduced]. It is also shown that the cutoff frequency of the WOTF is reduced from 2 to $2 - \Delta s$  with the increase of the annulus width.

\begin{figure}[htb!]
\centering
\includegraphics[width=0.95\linewidth]{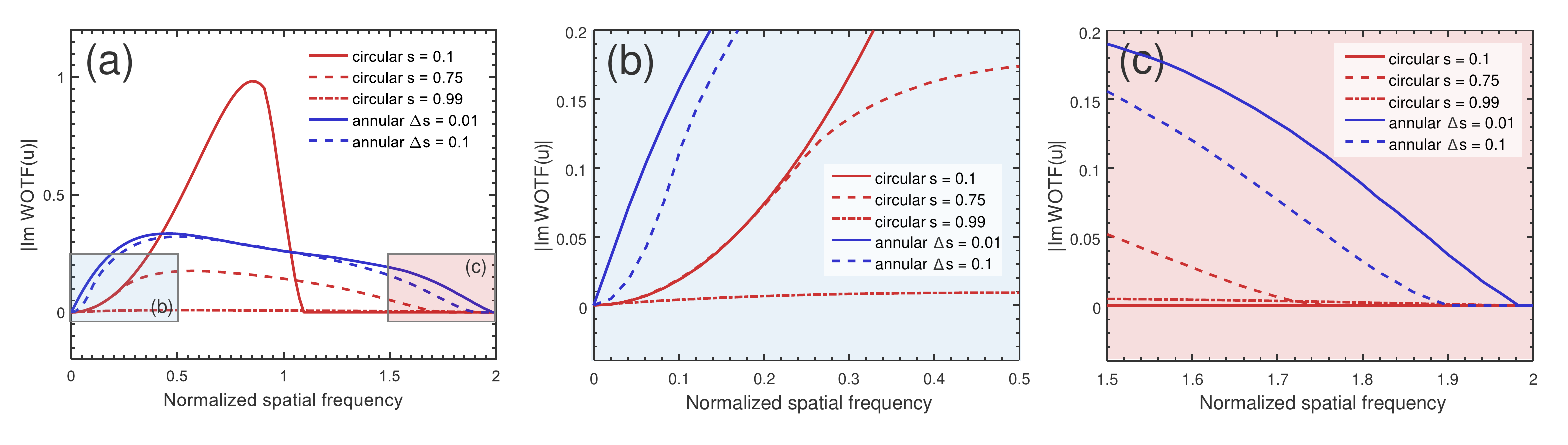}
\caption{(a) Magnitudes of the phase WOTFs of the annular illuminations ( $s = 0.01$ and $s = 0.1$) and circular illuminations (s = 0.1, 0.75, 0.99) when the defocus distance is 0.5 $\mu$m. For clarity, the two blue- and red-boxed regions are further enlarged in (b) and (c), respective.}
\label{fig6}
\end{figure}

From the results shown in Figs. \ref{fig4} and \ref{fig5}, it can be deduced that we should choose the diameter of the annulus to be equal to that of the objective pupil, and make its thickness as small as possible to optimize both phase contrast and imaging resolution. In this work, we use ${s_2} = 1$  , $\Delta \emph{s} = 0.1$ to ensure sufficient light throughput as well as a smooth WOTF response over board frequency range (up to $1.9N{A_{obj}}$ ). Figure \ref{fig6} further compares the magnitudes of the phase WOTFs of the annular illuminations ( $\Delta \emph{s} = 0.01$ and $\Delta \emph{s} = 0.1$ ) and circular illuminations (s = 0.1, 0.75, 0.99) under weak defocus (0.5 $\mu$m). The properties of the WOTF for the matched annular illumination are summarized as follows:

(1) The annular illumination WOTF provides a spatial frequency cut-off near the diffraction limit of an incoherent image system ($\sim$ $2N{A_{obj}}$, $1.99N{A_{obj}}$ for $\Delta \emph{s} = 0.01$, $1.9N{A_{obj}}$  for $\Delta \emph{s} = 0.1$) , but with additional imaginary component with strong response that allows for imaging of the phase information.

(2)	Compared with a conventional microscope with a wide-open circular aperture ( $s = 0.75$), the total phase contrast (the area enclosed by the WOTF curve and the frequency-axis) provided by the annular illumination is more than doubled (2.35 times for $\Delta \emph{s} = 0.01$, 2.11 times for $\Delta \emph{s} = 0.1$ ). Not only the frequency coverage is extended, but the response in both low and high spatial frequencies is significantly enhanced. For example, the relative gain of phase contrast at ${\bf{\bar u}} = 1.6$  is 4.36 times for $\Delta \emph{s} = 0.1$  and 5.82 times for $\Delta \emph{s} = 0.01$. The gain at ${\bf{\bar u}} = 1.7$ is even higher: 11.04 times for $\Delta \emph{s} = 0.1$  and 18.76 times for $\Delta \emph{s} = 0.01$ .

(3)	The total phase contrast provided by the annular illumination is comparable to that of a conventional microscope with nearly coherent ($s = 0.1$) illumination (91\% for $\Delta \emph{s} = 0.01$ , 82\% for $\Delta \emph{s} = 0.1$ ), but the response is much smoother and more extensive. The spatial frequency cut-off is almost doubled, and the phase contrast of low spatial frequencies is significantly increased. The relative gain of phase contrast at ${\bf{\bar u}} = 0.05$  is 5.79 times for $s = 0.1$  and 8.02 times for $s = 0.01$. The maximum gain is in the limit of zero frequency.

The above results demonstrate that replacing the conventional circular aperture with an annular one provides a convenient way to optimize the WOTF for achieving a broadband frequency coverage and enhanced response in both low- and high-frequencies. It is expected to achieve high-quality phase reconstruction and overcome the noise-resolution tradeoff in conventional TIE imaging by using only three intensity measurements. Moreover, the resulting phase WOTF contains no deep dips and zero-crossings in its pass-band, which removes the ill-posedness of the WOTF inversion. It is also worth mentioning that although these conclusions are drawn from one specific case ( $N{A_{obj}}{\rm{ = }}$0.8, $\lambda  = 550$nm , $\Delta z = 0.5$ $\mu$m), they can be generalized to different imaging parameters when defocus is small, \emph{i.e.}, the Fresnel number is large compared with unity (near-Fresnel region).

\subsection*{Experimental setup}

\begin{figure}[!t]
\centering
\includegraphics[width=0.95\linewidth]{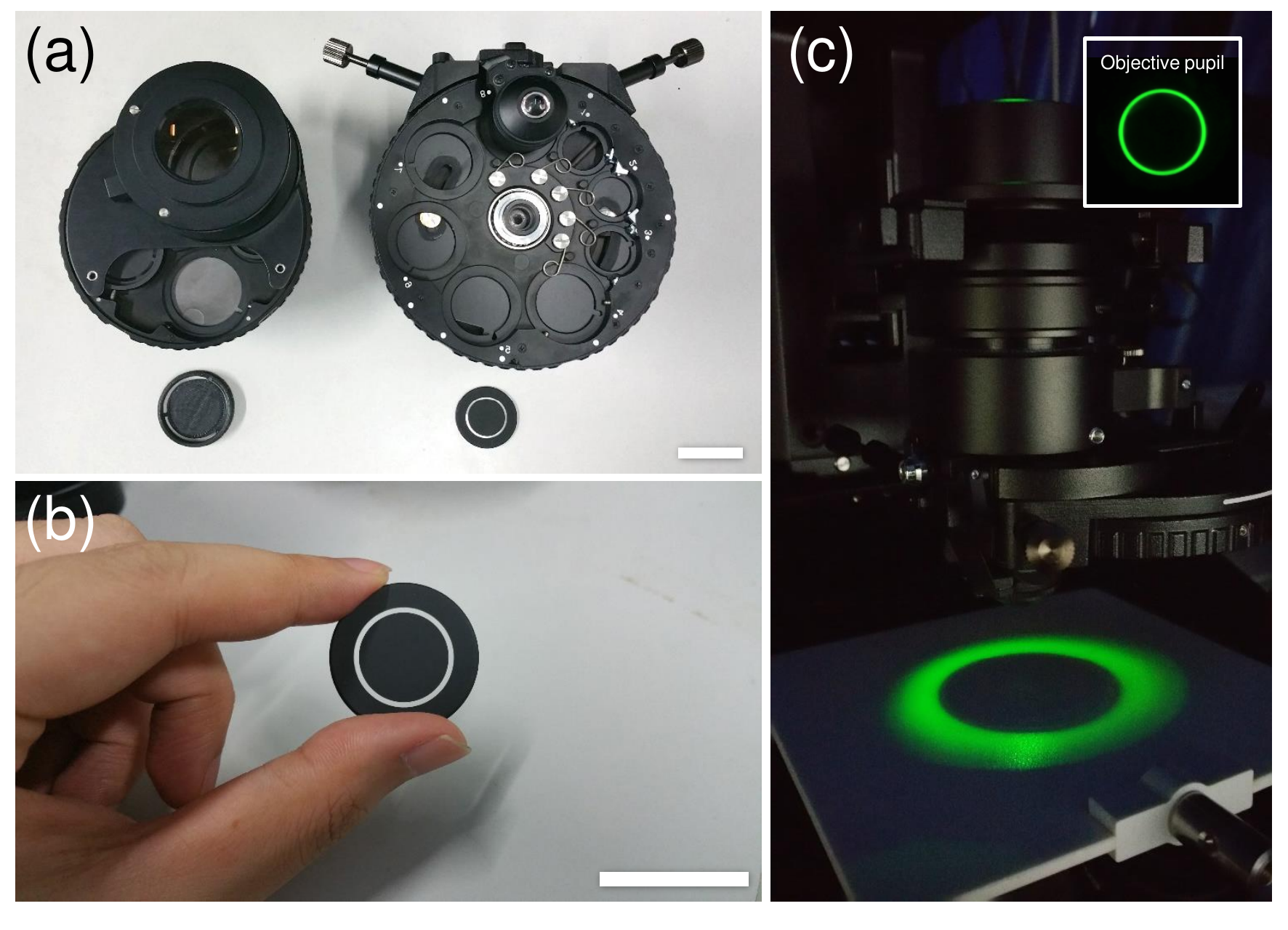}
\caption{Photographs of two different condensers used in the experiments. (a) IX2-MLWCD (left) and U-UCD8-2 (right) condenser with the corresponding annuli. (b) Annulus for the U-UCD8-2 condenser. Scale bars in (a) and (b) are both 30 mm. (c) The annular illumination generated from the U-UCD8-2 condenser with a dry type top lens. The inset shows the corresponding rear focal plane image of the objective.}
\label{fig7}
\end{figure}

\noindent
Experiments are performed based on a commercial inverted microscope (IX83, Olympus) equipped with a motorized focus drive with a minimum step size of 10 nm. The illumination is provided by the built-in a halogen lamp, filtered using a green interference filter [central wavelength $\lambda  = 550$nm, 10nm full width half maximum (FWHM) bandwidth] to provide quasi-monochromatic illumination. Two different types of condensers are used in our experiments (IX2-MLWCD and U-UCD8-2, Olympus), as shown in Fig. \ref{fig7}. The former one [Fig. \ref{fig7}(a) left] has a maximum NA of 0.55 and a long working distance of 27 mm. It accommodates our incubation chamber (INUF-IX3W-F1, Tokai Hit) for time-lapse imaging live cells, in cooperation with a 20$\times$, 0.4NA objective (PLN20$\times$, Olympus). The incubation chamber maintains an internal environment of 37$^\circ$C and provides humidified air with 5\% carbon dioxide. Besides, the IX2-MLWCD condenser is also used for the measurement of microlens array, with a 10$\times$, 0.4 NA objective (UPLSAPO10$\times$, Olympus). The U-UCD8-2 condenser [Fig. \ref{fig7}(a) right] is used for high-resolution imaging, with a maximum NA of 0.9 for dry type top lens (U-TLD, Olympus), and 1.4 for oil-immersion type top lens (U-TLO, Olympus). It is used for imaging the fixed human BMSC cells with the dry type top lens and a 40$\times$, 0.9NA objective (UPLSAPO40$\times$, Olympus), and the cheek cell with the oil-immersion type top lens and a 100$\times$, 1.4NA oil-immersion objective (UPLSAPO100$\times$, Olympus). The annular aperture used for the IX2-MLWCD is designed using SolidWorks and printed by a 3D printer (Replicator 2$\times$, Makerbot) in ``fine mode". The other annulus for the U-UCD8-2 condenser is made of a thin circular glass plate, with the opaque regions anodized and dyed with a flat-black pigment [Fig. \ref{fig7}(b)]. The width of the ring for both annular apertures is designed as 10 \% of the outer circle of the annulus. The two annuli are fitted into open slot positions in the respective condenser turret, and properly centered in the optical pathway. This is accomplished by examining the objective rear focal plane with a Bertrand lens inserted into one of the eyepiece observation tubes (in place of a normal eyepiece) to ensure that the annulus is properly inscribed inside the objective pupil [see the inset of Fig. \ref{fig7}(c)]. As a result, the sample is illuminated with a uniform light field, which has a cone-shell shaped angular distribution. Whenever annular apertures are utilized for phase imaging, the built-in diaphragm of the condenser should be opened to its widest position. For normal TIE phase imaging with a circular aperture, the microscope works in brightfield mode and the condenser diaphragm should be properly adjusted to achieve the desired coherent parameter \emph{s}.

\subsection*{Biological sample preparation}

\noindent
Human bone marrow stromal cells (BMSC) were cultured on coverslips in a 10 cm diameter of tissue culture dishes with medium composed of 15 \% (v/v) fetal bovine serum (FBS) , 100U/mL penicillin, 100 ug/mL streptomycin, 20 mg/L gentamicin, 1 ng/L fibroblast growth factor, and 3 g/L sodium bicarbonate in Alpha-MEM (Minimum Essential Medium Eagle - Alpha Modification)  media. At approximate 60\% confluence, cells were fixed on the coverslips with pre-warmed 1\% (v/v) glutaraldehyde solution in phosphate-buffered saline (PBS) for 10min. After washing three times with DI-water, the fixed cells are sealed between the coverslip and one microscope slide.

Buccal epithelial cells smear was collected from the cheek of a healthy adult volunteer, diluted in a drop of PBS, and spread out into a thin layer between a pair of 1.5-thick coverslips. The two coverslips were separated by an adhesive imaging spacer (Grace Bio-Labs SS1X9-SecureSeal Imaging Spacer: inner diameter, 9 mm; thickness, 0.12 mm; 18 mm $\times$ 18 mm) to avoid deforming the specimen.

HeLa cells were seeded (at an initial density of ~ 300 cells/cm$^2$) and in a 35 mm glass-bottom Petri dish in Dulbecco’s Modified Eagle’s Medium (DMEM) supplemented with 10\% FBS, and 1\% penicillin streptomycin. Cells were incubated at 37$^\text{o}$C in humidified atmosphere of 5 \% carbon dioxide for 8 hours to allow attachment. After that, cells were washed twice with PBS and pre-warmed fresh medium was added. Then, the cells were placed on the stage incubator of the microscope for long-term time-elapse imaging.

For cell drug treatment experiment, paclitaxel powder was dissolved in dimethyl sulfoxide (DMSO) to a concentration of 10 nM. HeLa cells were cultured in the above-mentioned medium and environment before they were exposed to paclitaxel. After 24 hours of incubation, the original medium was replaced by fresh medium plus paclitaxel at the stated concentration. The cells were then subjected to time-lapse imaging within the stage incubator of the microscope.

\subsection*{Phase reconstruction by AI-TIE }

\noindent
The AI-TIE reconstruction algorithm is similar with the conventional TIE and the WOTF inversion approaches described in Section \textbf{Quantitative phase reconstruction by inversion of TIE and WOTF}, which contains the following 3 steps:

(1)	Calculate the WOTF function $WOTF\left( {\bf{u}} \right)$ based on Eq. (\ref{11}) according to experimental parameters used in the corresponding configurations (e.g., NA of the objective, width of the annulus, defocus distance). Only the imagery part of the WOTF  ${\mathop{\rm Im}\nolimits} \left[ {WOTF\left( {\bf{u}} \right)} \right]$ is used for phase reconstruction.

(2)	Capture one in-focus image ${I_0}\left( {\bf{x}} \right)$ and two slightly defocused images with equal and opposite defocus distances $\pm \Delta z$ (${I_{ \pm \Delta {\rm{z}}}}\left( {\bf{x}} \right)$). The defocus distance $\Delta z$  should be carefully chosen in order to produce a well-behaved WOTF function (the Fresnel number is large compared with unity).

(3)	Reconstruct the phase by Fourier space de-convolution based on Eq. (\ref{22}). Since the resulting phase WOTF of AI-TIE contains no deep dips and zero-crossings in its pass-band, there is no need for regularization. The regularization parameter $\alpha$ in Eq. (\ref{22}) can be set to an infinitely small value (e.g., variable "eps" in MATLAB).

\subsection*{Simulation of phase-contrast and differential-interference-contrast (DIC) images}

\noindent
Once the complex field of the sample is obtained by phase retrieval, it is possible to emulate other forms of phase contrast microscopy computationally \cite{84,85} (although the phase and amplitude information does not fully characterize a partially coherent field, it does allows other imaging modalities to be easily simulated). The differential-interference-contrast (DIC) image is obtained by taking the phase gradient in the direction of the image shear (45$^\circ$ as in our experiments): ${I_{DIC}}\left( {\bf{x}} \right) = 2{I_0}\left( {\bf{x}} \right)\left[ {1 + \delta {\bf{x}} \cdot \nabla \phi \left( {\bf{x}} \right)} \right]$ , where $\delta {\bf{x}} = \left( {1/\sqrt 2 ,1/\sqrt 2 } \right)$. For phase contrast images, after taking the Fourier transform, a phase shift of $\pi /2$ is digitally applied to the low frequency component (${\kern 1pt} \left| {{\bf{\bar u}}} \right| \le 0.1$) of the object complex field in the Fourier space. The modified complex field is then followed by an inverse Fourier transform to create a digital phase contrast image of the specimen.

\subsection*{Cell area and dry mass calculation}

\noindent
To perform area and dry mass measurements on cells, each cell must be segmented from the background and individualized from other cells. We first separate the pixels corresponding to cell regions from the pixels of the background, which is realized by an automatic thresholding performed on the phase image, followed by a binary morphologic dilatation with a 7 $\times$ 7 disk-shaped structuring element. After masking out the background, the local maxima of the 25 $\times$ 25 Gaussian-smoothed phase image are used to detect the approximate location of each cell. These local maxima are also used as initial seed points of a watershed algorithm to deduce the precise contour of each cell.

After cell segmentation, the area of each individual cell can be easily obtained by counting the number of segmented pixels. The area summation of the segmented cells is used to calculate the confluence of the cell culture, which is the proportion of the total examined field (332.8 $\times$ 332.8 $\mu$m$^2$) covered by cells. The dry mass for each cell is calculated as the sum of the dry mass density integrated over all cell area. The relation between the dry mass density and phase is given by $\rho \left( {\bf{x}} \right) = \frac{\lambda }{{2\pi \gamma }}\phi \left( {\bf{x}} \right)$, where $\gamma {\rm{ = }}0.2ml/g$  is known as the refractive increment, which relates the change in concentration of protein to the change in refractive index \cite{86}.

\section*{RESULTS AND DISCUSSION}

\subsection*{Comparison between annular illumination TIE and circular illumination TIE}

\begin{figure}[htb!]
\centering
\includegraphics[width=1.0\linewidth]{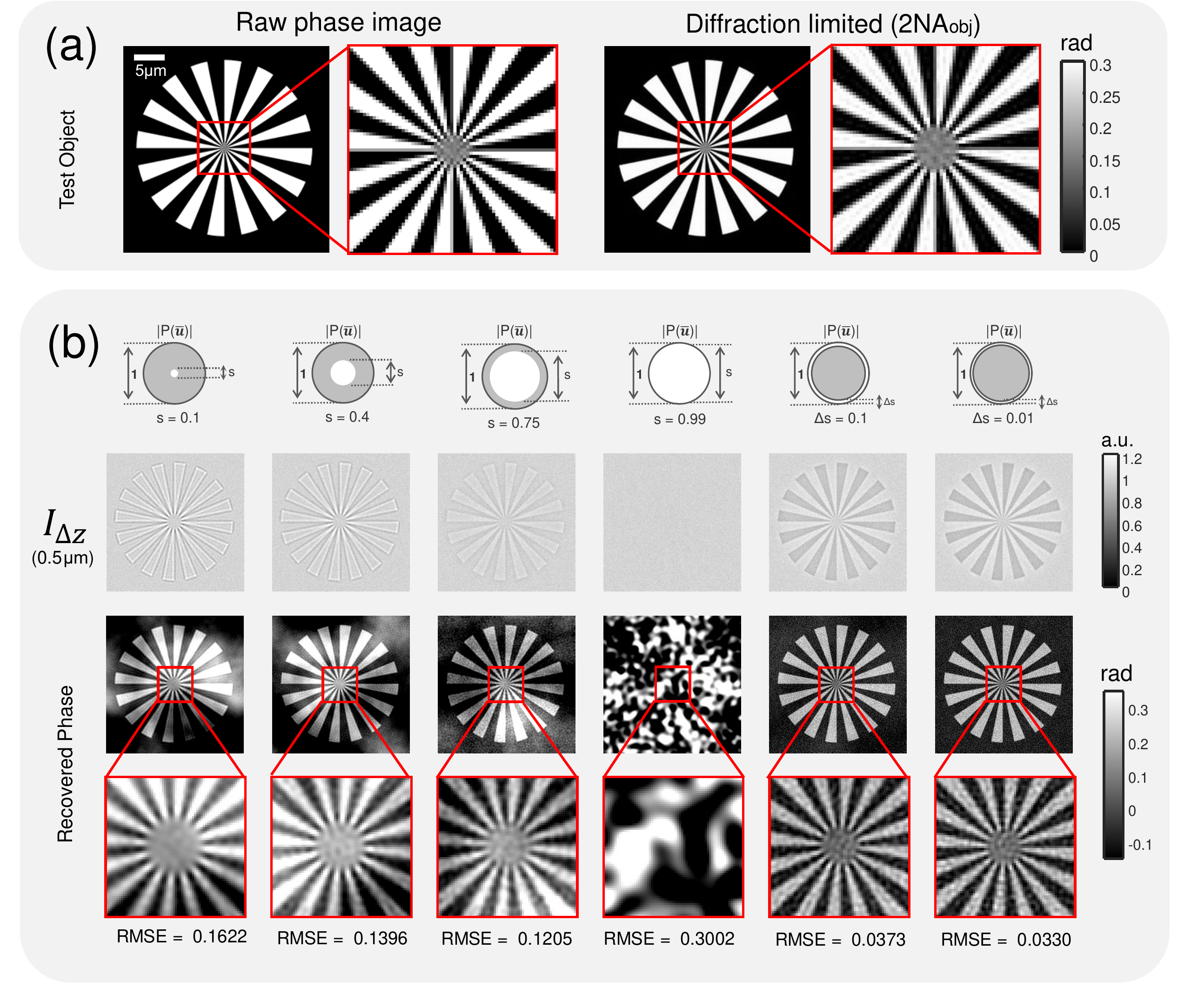}
\caption{Comparison between annular illumination TIE and circular illumination TIE. (a) The raw Siemens star image and the corresponding best diffraction limited image can be achieved based on the simulation parameter. (b) Comparison of over-defocus images and reconstruction results of different illumination settings for a small defocus distance ($\Delta$z = 0.5 $\mu$m).}
\label{fig8}
\end{figure}

\noindent
The proposed AI-TIE is first compared with the traditional circular illumination TIE based on simulations. The Siemens star image \cite{87} is used as an example phase object [shown in Fig. \ref{fig8}(a)] which is defined on a grid with 256 $\times$ 256 pixels with a pixel size of 0.13 $\mu$ m $\times$ 0.13 $\mu$m. The wavelength of the illumination is 550 nm, and the $N{A_{obj}}$ is 0.80. For such an imaging configuration, the best phase imaging resolution can be achieved is  , which is also shown in Fig. \ref{fig8}(a). For partially coherent image calculation, we use the Abbe’s method, in which each sub-image corresponding to point source in the condenser aperture plane is superimposed at the image plane according to Eq. (\ref{1}). Defocus is modeled as the angular spectrum function applied at the pupil plane of the objective [Eq. (\ref{10})]. To simulate the noise effect, each defocused image is corrupted by Gaussian noise with standard deviation of 0.01.

\begin{figure}[htb!]
\centering
\includegraphics[width=0.8\linewidth]{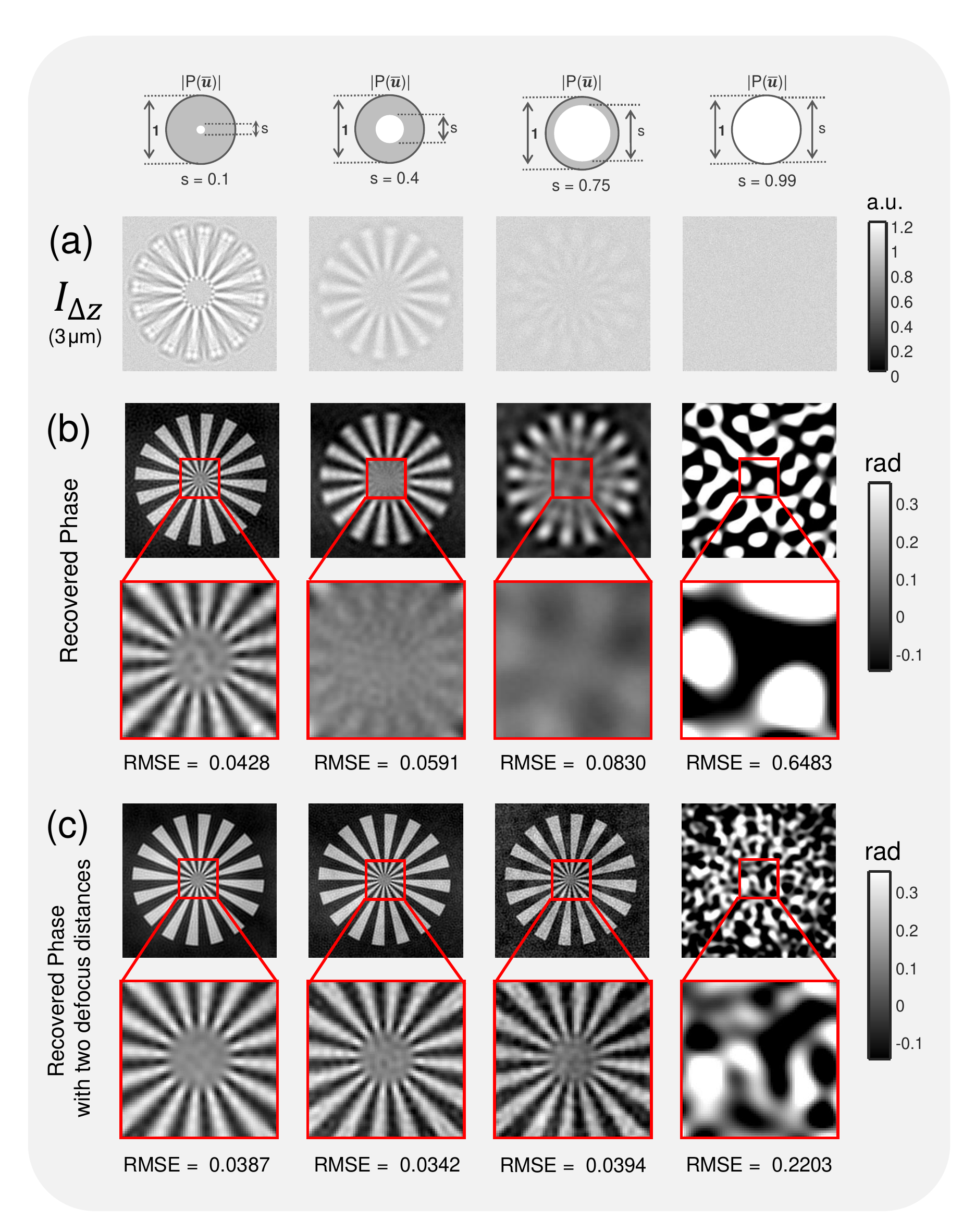}
\caption{Comparison of over-defocus images (a) and reconstruction results of different illumination settings (b) for a large defocus distance ($\Delta$z = 3 $\mu$m) . The synthesized phases using both small ($\Delta$z = 0.5 $\mu$m) and large defocus distances ($\Delta$z = 3 $\mu$m) are shown in (c).}
\label{fig9}
\end{figure}

In Fig. \ref{fig8}(b), we compare the defocused images and the phase retrieval results of different illumination settings for a small defocus distance ($\Delta$z = 0.5 $\mu$m). The metric used to measure the accuracy of phase retrieval is given by the root mean square error (RMSE), which quantifies the overall difference between the ideal phase and the retrieved phase. For the case of circular illumination, the overall phase contrast reduces with the increase in coherent parameter \emph{s}, which is in coincidence with the WOTF analysis [Fig. \ref{fig3}]. The poor response at low-spatial frequencies leads to cloud-like artifacts superimposed on the reconstructed phases. Besides, the phase imaging resolution is improved by opening up the condenser diaphragm (increasing the coherent parameter \emph{s}). However, for the case of nearly matched illumination (\emph{s} = 0.99), the washout in phase contrast prevents any recognizable phase information to be reconstructed, leading to significant artifacts and a very large RMSE. Furthermore, the washout effect can be more prominent when the defocus distance increases to 3 $\mu$m, as shown in Fig. \ref{fig9}. By combining the two phase reconstructions at small and large defocus distances with the least-squares weighting method \cite{50,58}, the low-frequency noise is greatly reduced, as shown in Fig. \ref{fig9}(c). Without capturing additional images at a larger defocus distance, the phase contrast, especially for low-frequency components can be significantly enhanced by using AI-TIE. The Siemens star appears dark in the defocused image, demonstrating the negative phase contrast as predicted by the theory. The strong phase contrast is finally converted to the quantitative phase images by WOTF inversion, resulting in high-quality reconstructions with a uniform background and improved resolution. The RMSE values for the AI-TIE are comparable with the conventional two-distance TIE approaches and significantly lower than the case when only single defocus distance is used. Besides, the theoretical resolution for AI-TIE is improved to $1.9N{A_{obj}}$ for $\Delta \emph{s} = 0.1$ and $1.99N{A_{obj}}$ for $\Delta \emph{s} = 0.01$, which approaches to the incoherent limit [Fig. \ref{fig8}(a) right]. It is also demonstrated in \textbf{Supplementary Information C} that the annular illumination TIE is also quite robust against the model mismatch (mis-estimation of the WOTF) and the aberration presented in the imaging optics. In contrast, the frequency response of circular aperture includes deep dips and zero-crossings (especially when the defocus distance is large) so that a mis-estimation of the WOTF will lead to larger phase errors during the inversion (see\textbf{ Supplementary Information C} for details).

\subsection*{Imaging of fixed human BMSC cells}

\begin{figure}[!b]
\centering
\includegraphics[width=0.95\linewidth]{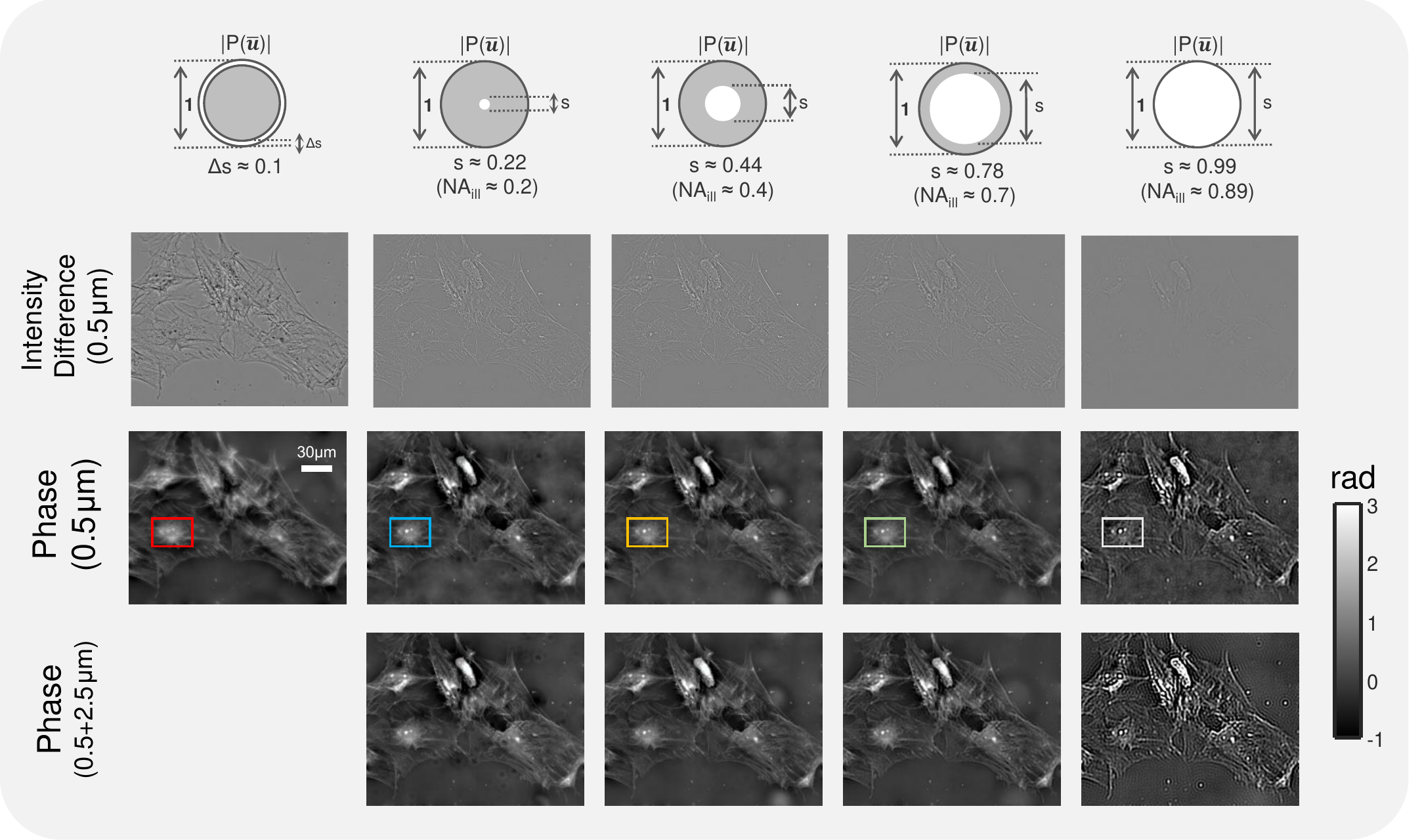}
\caption{Comparison of annular illumination TIE with circular illumination TIE for imaging of fixed human BMSC cells. First row: intensity difference from two defocused images with $\Delta$z = $ \pm $ 0.5 $\mu$m. Second row: phase reconstruction with of different illumination settings with single defocus distance ($\Delta$z = 0.5 $\mu$m). Second row: phase reconstruction with different illumination settings and single defocus distance. Third row: phase reconstruction with different illumination settings and two defocus distances ($\Delta$z = 0.5, 2.5 $\mu$m).}
\label{fig10}
\end{figure}

\begin{figure}[htb!]
\centering
\includegraphics[width=1.0\linewidth]{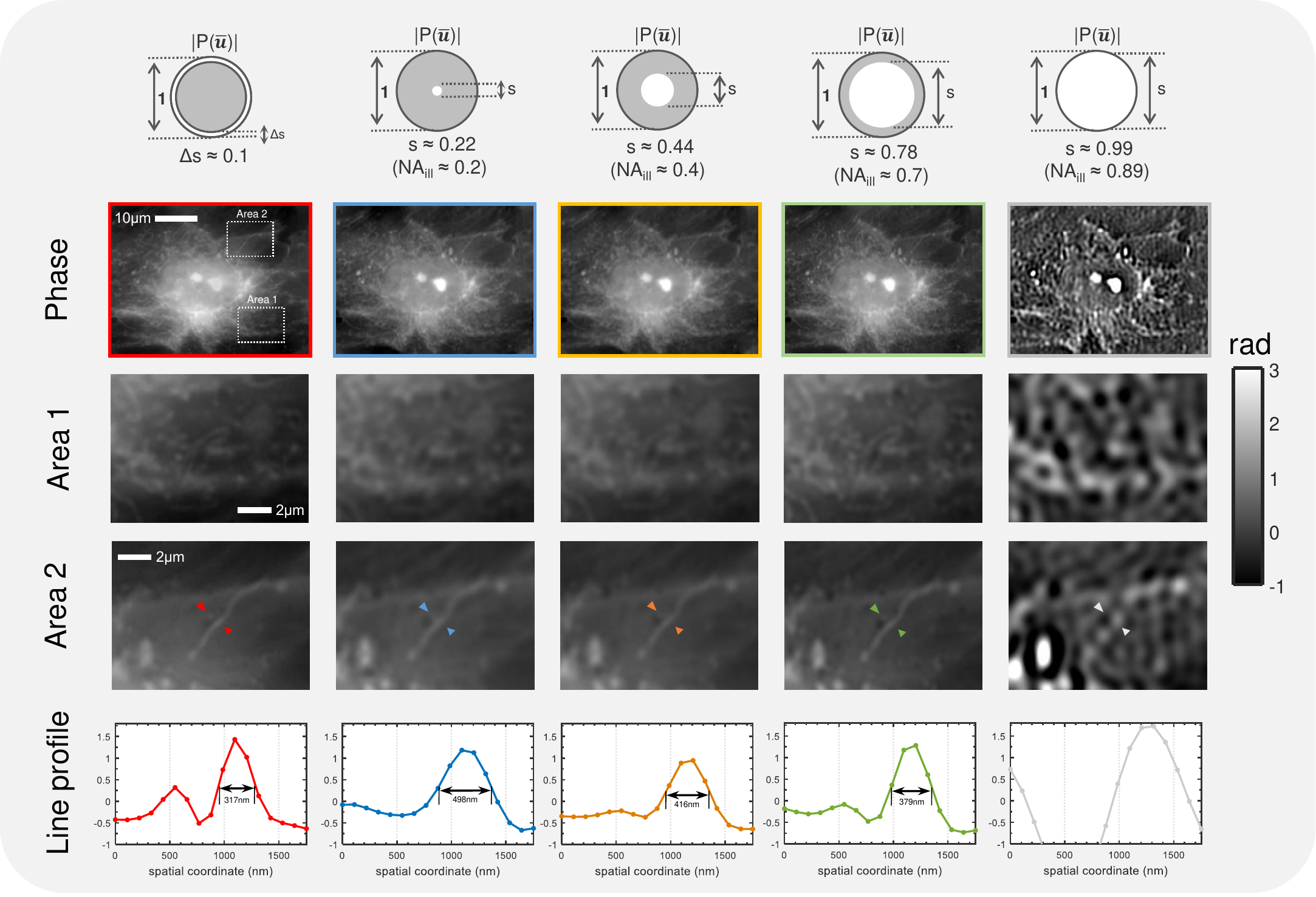}
\caption{Comparison of imaging resolution between annular illumination TIE and circular illumination TIE for imaging of fixed human BMSC cells. Two areas (boxed regions) containing intracellular organelles near the nucleus are enlarged and shown in the second and third rows. The phase line profiles along the respective arrow are shown in the bottom row.}
\label{fig11}
\end{figure}

\noindent
We then compare the proposed AI-TIE with the circular illumination TIE experimentally on fixed human BMSC cells. Here, we use U-UCD8-2 condenser with the dry type top lens (maximum NA of 0.9) and a 40$\times$, 0.9NA objective (UPLSAPO 40$\times$, Olympus) that provide a best possible lateral resolution of 306 nm (corresponding to an effective NA of 1.8). In this work, the theoretical resolution limit is considered in terms of \emph{minimum resolvable pitch for periodic structures} (wavelength divided by the effective NA), which is a conservative estimate compared to other criterions based on minimum distinguishable distance between two points (\emph{e.g.} underestimate the resolution limit by a factor of 1.47 compared to the coherent Sparrow criterion \cite{88}). The defocused images and phase retrieval results of different illumination settings for a small defocus distance ($\Delta$z = 0.5 $\mu$m) are compared in Fig. \ref{fig10}. To illustrate the imaging resolution more clearly, a small region near the cell nucleus (boxed area) for each phase image is further magnified in Fig. \ref{fig11}, with two zoom-ins showing intracellular organelles near the nucleus at a submicrometer-scale. The results shown in Figs. \ref{fig10} and \ref{fig11} are generally consistent with the theoretical estimate and simulation results. For the case of circular apertures, using a small defocus distance leads to low-frequency noise, which needs to be suppressed by combining additional intensity measurements at a larger defocus distance ($\Delta$z = 2.5 $\mu$m). For large \emph{s} = 0.99, the reconstruction is strongly corrupted by the artifacts due to the low response of the WOTF. The annular illumination provided improved and inverse phase contrast, resulting in a low-noise phase reconstruction by using only a single defocus distance. In Fig. \ref{fig11}, it can be clearly seen that the small point (transported vesicle) and line structure (Golgi apparatus) are clearly resolved in the reconstruction of AI-TIE. The FWHM of the Golgi apparatus along the arrow is measured to be 317 nm, which agrees with the expected resolution (321 nm, 1.71 NA). While these detailed subcellular structures cannot be readily observed in the reconstruction results of the circular illuminations when \emph{s} = 0.22 and \emph{s} = 0.44. When \emph{s} increases to 0.7, the structure can be slightly resolved, with a FWHM of 379 nm, which is a bit larger than the theoretical value (359 nm, 1.53 NA). The difference may result from the optical aberration or the inaccuracy in estimating the coherent parameter.

In this experiment, the phase range of cells is over 3 rad. Thus, they cannot be considered as weak phase object (which typically requires the total phase shift introduced by the object to be less than $\pi /2$ \cite{74}). Though this weak object assumption (in both amplitude and phase) is essential for the derivation of the WOTF, we demonstrate in both simulation and experiment that the validity of AI-TIE does can extend far beyond this range due to the fact that other types of linearization conditions, like the weak defocusing and slowly varying phase approximation can be satisfied implicitly \cite{48,58} (see \textbf{Supplementary Information D} and Fig. \ref{fig13} for details).

\subsection*{High resolution imaging of buccal epithelial cells}

 \begin{figure}[!htp]
\centering
\includegraphics[width=0.8\linewidth]{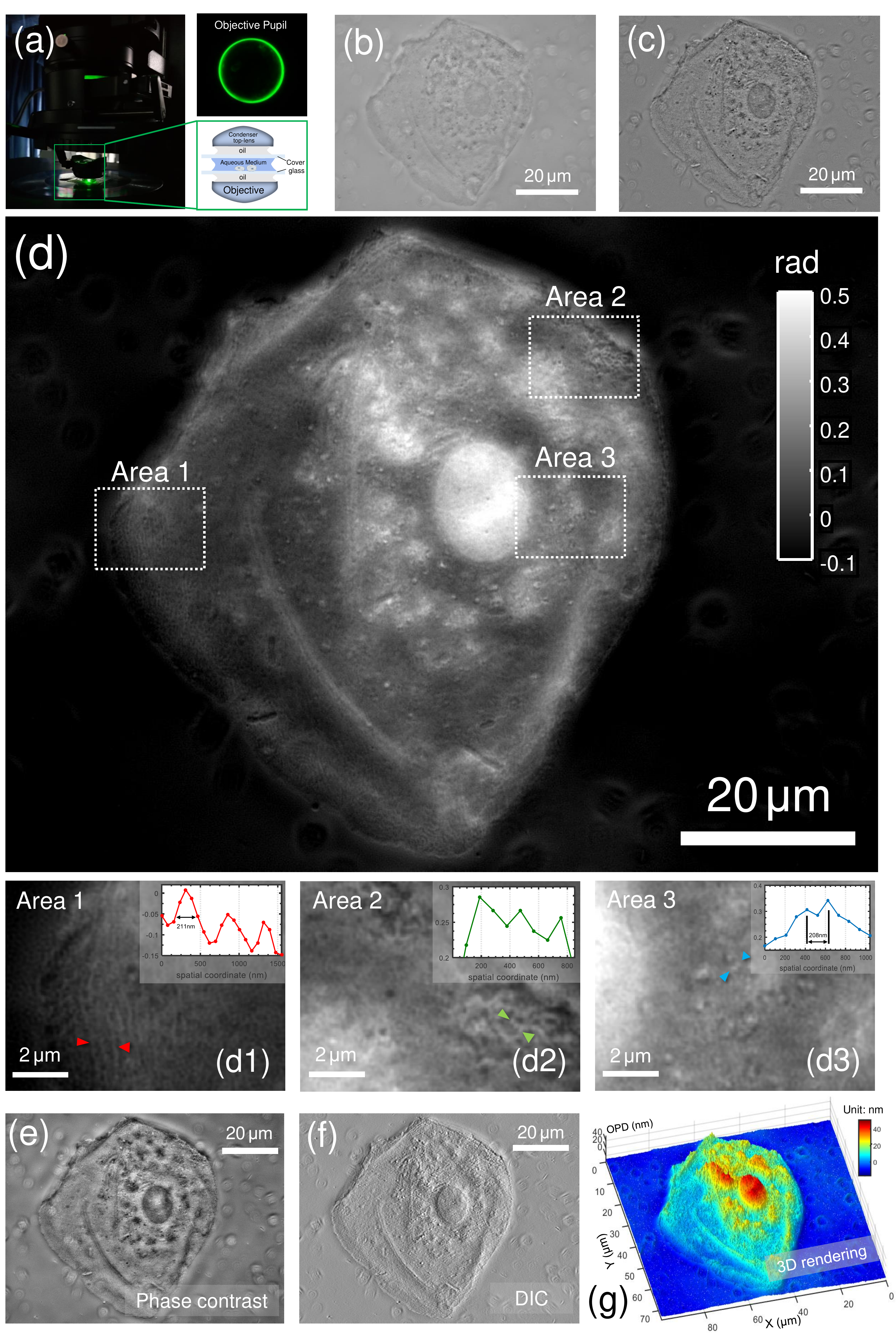}
\caption{High-resolution imaging of buccal epithelial cell. (a) Optical configuration with the corresponding rear focal plane image of the objective. (b) In-focus intensity image. (c) Intensity difference image. (d) Reconstructed phase image. Three areas of interest (boxed regions) are enlarged in (d1)-(d3). The insets shows line profiles taken at different positions in the cell. (e) Simulated phase-contrast image; (f) Simulated DIC image; (g) Pseudo-color 3D rendering of the cell optical length.}
\label{fig12}
\end{figure}

\noindent
Human buccal epithelial cells (cheek cells) sandwiched between a couple of microscope coverslips [Fig. \ref{fig12}(a)] are imaged with U-UCD8-2 condenser with the oil-immersion type top lens (maximum NA of 1.4) and a high NA = 1.4 oil-immersion objective (UPLSAPO100$\times$, Olympus). Though cheek cells are excellent transparent specimens especially when the microscope is perfectly in-focus [Fig. \ref{fig12}(b)], in the annular illumination configuration, the phase variation produces strong phase contrast with slight defocusing ($ \pm $ 250 nm), generating a high-contrast intensity difference images revealing much clearer information about the actual structure of the specimen [Fig. \ref{fig12}(c)]. The quantitative phase image reconstructed by AI-TIE is shown in Fig. \ref{fig12}(d), with three zoomed areas within the entire cell region depicted in Figs. \ref{fig12}(d1)-(d3). There, the optically thick nucleus, squamous structure on the cell membrane, and some cytoplasmic organelles are shown with high contrast and clarity. The insets of Figs. \ref{fig12}(d1)-(d3) show line profiles taken at different positions in the cheek cell. The regular fenestrations on the cell membrane with $\sim$ 219 nm FWHM of the structure size are clearly revealed in Fig. \ref{fig12}(d1). In Fig. \ref{fig12}(d3), the smallest yet clearly identified spherical structures have a separation of only 208 nm [Fig. \ref{fig12}(d4)]. These results confirm that the resolution of the phase image is at least on the order of 208 nm, which corresponds to an effective NA of 2.66 ($1.9N{A_{obj}}$ at 550 nm wavelength) for our setup. To our knowledge, such high-resolution QPI capability of TIE has never been demonstrated up to now. Once quantitative phase information is obtained, other types of microscopic modalities such as phase contrast, and DIC can be numerically simulated without using actual hardware, as demonstrated in Figs. \ref{fig12}(e) and \ref{fig12}(f). The phase contrast image highlights the differences of refractive index and thickness of microscopic structures, with typical bright `halo' surrounding the cell periphery. The DIC image is calculated from the phase gradient in the direction of the image shear (45$^\circ$), which produces a high-contrast pseudo-relief imaging effects that effectively enhance the gradient of optical paths for both low and high spatial frequency details. The quantitative phase profile is plotted again in Fig \ref{fig12}(g) as a surface plot, where the $z$-axis corresponds to the computed optical thickness. The pseudo-color 3D rendering of the specimen reveals high-quality and high-contrast surface details on the cell and provides an accurate profile of optical thickness. It should be mentioned that the optical thickness distribution (or optical path length) shown in Fig. \ref{fig12}(g) is calculated directly from the obtained phase $OPD\left( {\bf{x}} \right) = \frac{\lambda }{{2\pi }}\phi \left( {\bf{x}} \right)$ , which is an accumulation of refractive index over the cellular thickness. If one wants to obtain the true (physical) thickness of the cell, additional decoupling procedure by varying the refractive index of the surrounding medium \cite{89} or more complicated tomography techniques \cite{19,90,91,96} need to be applied.

\subsection*{Characterization of a microlens array}

\begin{figure}[!b]
\centering
\includegraphics[width=0.85\linewidth]{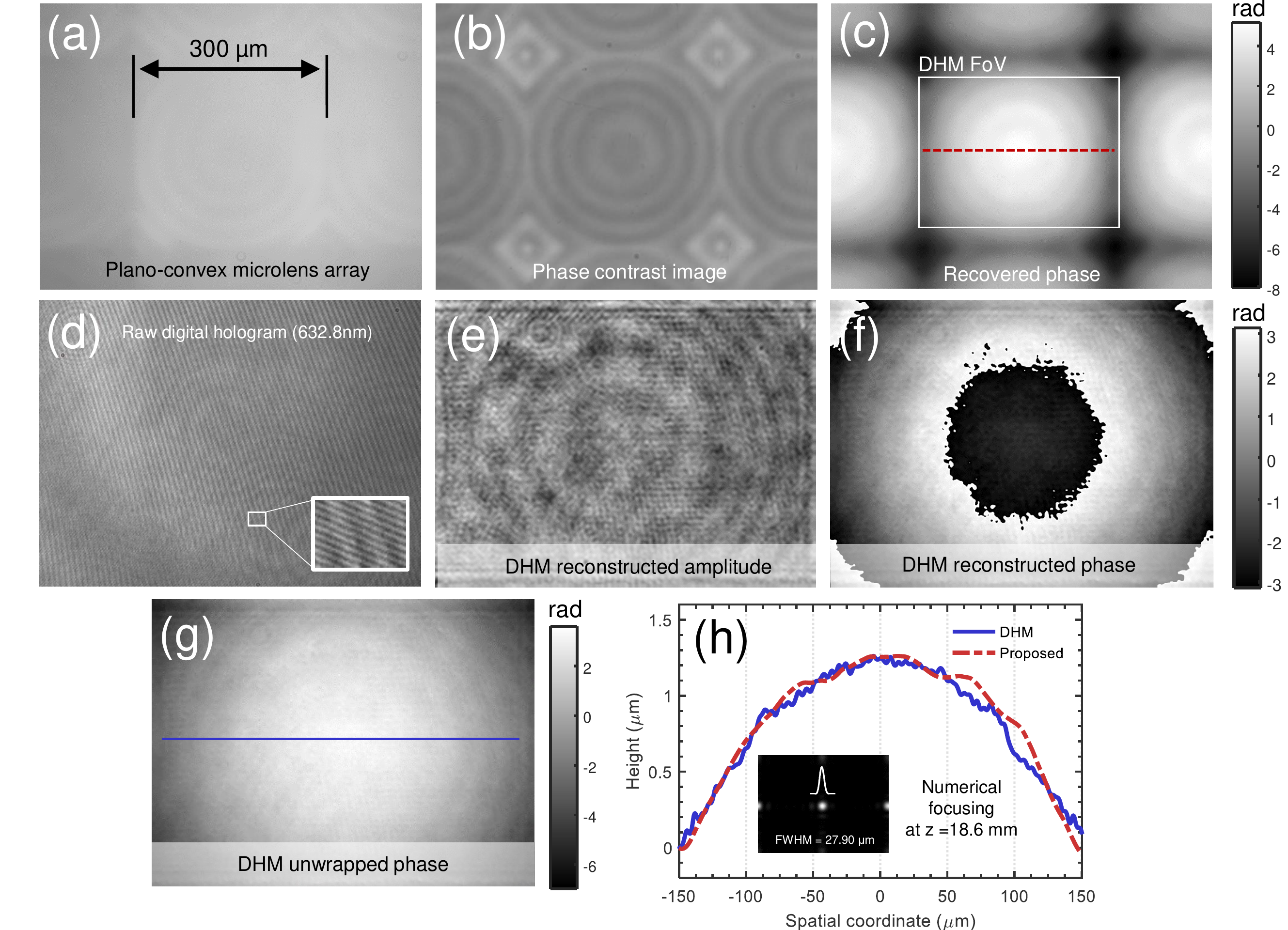}
\caption{Microlens array characterization using AI-TIE. (a) In-focus intensity image. (b) Phase contrast microscope image. (c) Reconstructed phase image by AI-TIE. (d) Digital hologram captured by a DHM system. The carrier fringes can be easily seen in the magnified area. (e) DHM reconstructed amplitude. (f) DHM reconstructed (wrapped) phase. (g) Unwrapped phase. (h) Height line profiles corresponding to the red dashed line in (c) and the blue solid line in (g), respectively.}
\label{fig13}
\end{figure}

\noindent
To demonstrate the accuracy of the phase reconstruction for AI-TIE, a plano-convex microlens array (MLA300-7AR, Thorlabs) is measured with the IX2-MLWCD condenser and a 10$\times$, 0.4NA objective (UPLSAPO10$\times$, Olympus). This microlens array is specifically designed for Shack-Hartmann sensor applications, with a pitch of 300 $\mu$m (square) and a focal length of 18.6 mm. Figure \ref{fig13}(c) shows the quantitative phase image reconstructed by annular illumination based TIE using the in-focus image [Fig. \ref{fig13}(a)] and two defocused images at $ \pm $ 1 $\mu$m. Note that the phase map without 2$\pi$ discontinuities is directly obtained without phase unwrapping. Besides, some concentric zone rings are clearly reflected in the phase map, suggesting the lens profile is not perfectly circular. This observation is further supported by the phase contrast [Fig. \ref{fig13}(b)] image taken with a 10$\times$, 0.3NA phase-contrast objective (UPLANFI10$\times$PH, Olympus), in which the ring structure is clearly revealed as well. To assess the accuracy of the phase measurement, the same sample is also measured using a digital holographic microscope (DHM) system equipped with a 60$\times$, 0.85 NA microscope objective (laser wavelength 632.8 nm). Figure \ref{fig13}(d) shows the raw digital hologram captured by the system. The high-frequency carrier fringes due to the off-axis geometry are clearly visible in the enlarged area. Its numerically reconstructed amplitude and wrapped phase are shown in Fig \ref{fig13}(e) and \ref{fig13}(f), respectively. Albeit a bit noisy due to the laser speckle, the underlying ring structure is still perceivable in both the amplitude and phase reconstructions. After phase unwrapping, the continuous phase map [Fig. \ref{fig13}(g)] can be converted to the phase to the physical thickness of the lens [the refractive index of lens material (fused silica) is 1.46 at 550 nm, and 1.457 at 632.8 nm]. Thickness profiles for the same lens from the array taken along the red dashed line in Fig. \ref{fig13}(c) and the blue solid line in Fig. \ref{fig13}(g) are compared quantitatively in Fig. \ref{fig13}(h), showing a reasonable agreement. The small discrepancy may be attributed to the coherent noise effect or inaccuracies in determining the "best focus" plane in the DHM reconstruction. Combining the measured intensity and reconstructed phase map from our approach, the complex field can be numerically propagated to the known focal plane of the lens (18.6 mm from the input plane), resulting a sharp intensity focus [see the inset of Fig. \ref{fig13}(h)]. Due to the imperfection of the lens profile, the FWHM of the intensity focus is measured to be 27.90 $\mu$m, which is slightly larger than the ideal airy disk of a perfect lens with the same parameters (20.96 $\mu$m).

\subsection*{Long-term time-lapse imaging of HeLa cell dividing in culture}

\begin{figure}[!b]
\centering
\includegraphics[width=1.0\linewidth]{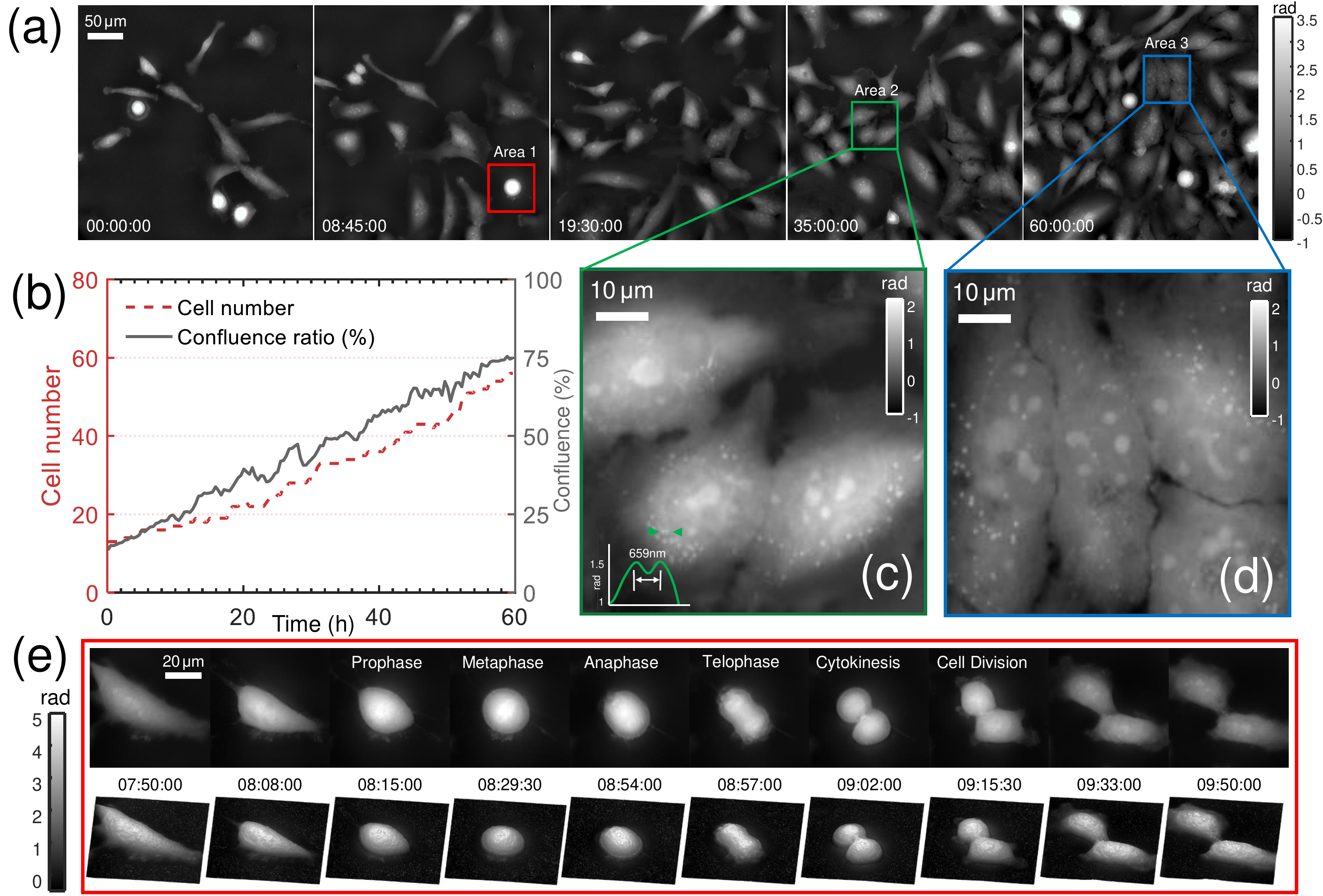}
\caption{Time-lapse phase imaging of HeLa cell division over a long period (60 h). (a) Representative quantitative phase images at different time points. (b) The change of cell number and confluence ratio over the culture passage period. (c) and (d) show the magnified views corresponding to two regions of interest  (Area 2 and Area 3). (e) 10 selected time-lapse phase images and the corresponding 3D renderings showing the morphological features of a dividing cell (Area 1) at different stages of mitosis.}
\label{fig14}
\end{figure}

\noindent
The high-resolution QPI capability of the proposed AI-TIE provides unique possibilities for the label-free imaging of cell growth in culture, using repeated imaging of cultures to assess the progression towards confluence over designated periods of time. In Fig. \ref{fig14}, we show five representative quantitative phase images of the human cervical adenocarcinoma epithelial (HeLa) cell division process over the course of 60 h, measured with the IX2-MLWCD condenser and a 20$\times$, 0.4NA objective (PLN20$\times$, Olympus). A time-lapse movie created with one phase reconstruction per 30 second is provided in \textbf{Supplementary Video 1}. The low light exposure and the absence of the fluorescent agents eliminate any concerns about the phototoxicity and photobleaching. Besides, the high degree of contrast achieved in the phase maps makes them especially amenable to segmentation of each individual cells. The area summation of the segmented cells provides a measure of confluence of the cell culture, expressed as a percentage of the total field examined. As shown in Fig. \ref{14}(b), an approximately linear growth rate was observed over the 60 h period with the confluence ratio increasing from ~13\% at 0 h to ~76\% after 60 h. It can be also found that the growth curve of cell confluence is closely correlated with the cell number, producing a correlation coefficient of  $r^2$ = 0.9654. In the two zoom-ins of the phase images shown in Figs. \ref{fig14}(c) and \ref{fig14}(d), subcellular features, such as cytoplasmic vesicles, are clearly observed. The line profile in Fig. \ref{fig14}(c) demonstrates a valley between two closely spaced features with center-to-center distance of 659 nm, indicating the phase imaging resolution is better than this value [the theoretical resolution (723 nm, 0.76 NA) is clearly a underestimate]. Besides, the temporal phase fluctuation (standard deviation) is measured to be only 0.007 rad (by performing 100 independent measurements without the presence of the sample), suggesting a good repeatability of the phase measurement. In Fig. \ref{fig14}(e), we further selected one cell [corresponding to the red-boxed region shown in Fig. \ref{fig14}(a)] to study its morphology during division, which spanned over about 2 hours (see \textbf{Supplementary Video 2} for the full 60-hour time-lapse 3D movie). The high resolution phase images clearly reveals the cell morphology during different phases of mitosis. Prior to the division, the mother cell rounded up and organized its cremation. The ``white bar'' on top of the cell at time point 8 h 29 min corresponds to the chromatin just before the separation of the two sets of genes. At time point 8 h 56 min, the chromosome divided into two sets and moved to opposite sides of the cell. Then the cytoplasm divided and cell wall pinched off at 9 h 02 min. Finally, two individual daughter cells are individualized and spread out. These results demonstrate that AI-TIE is capable of imaging unlabeled cells in the traditional environment of an inverted microscope, allowing for high resolution QPI over an extended period of time.

\subsection*{Multi-modal computational imaging of HeLa cell apoptosis}

\begin{figure}[!t]
\centering
\includegraphics[width=0.95\linewidth]{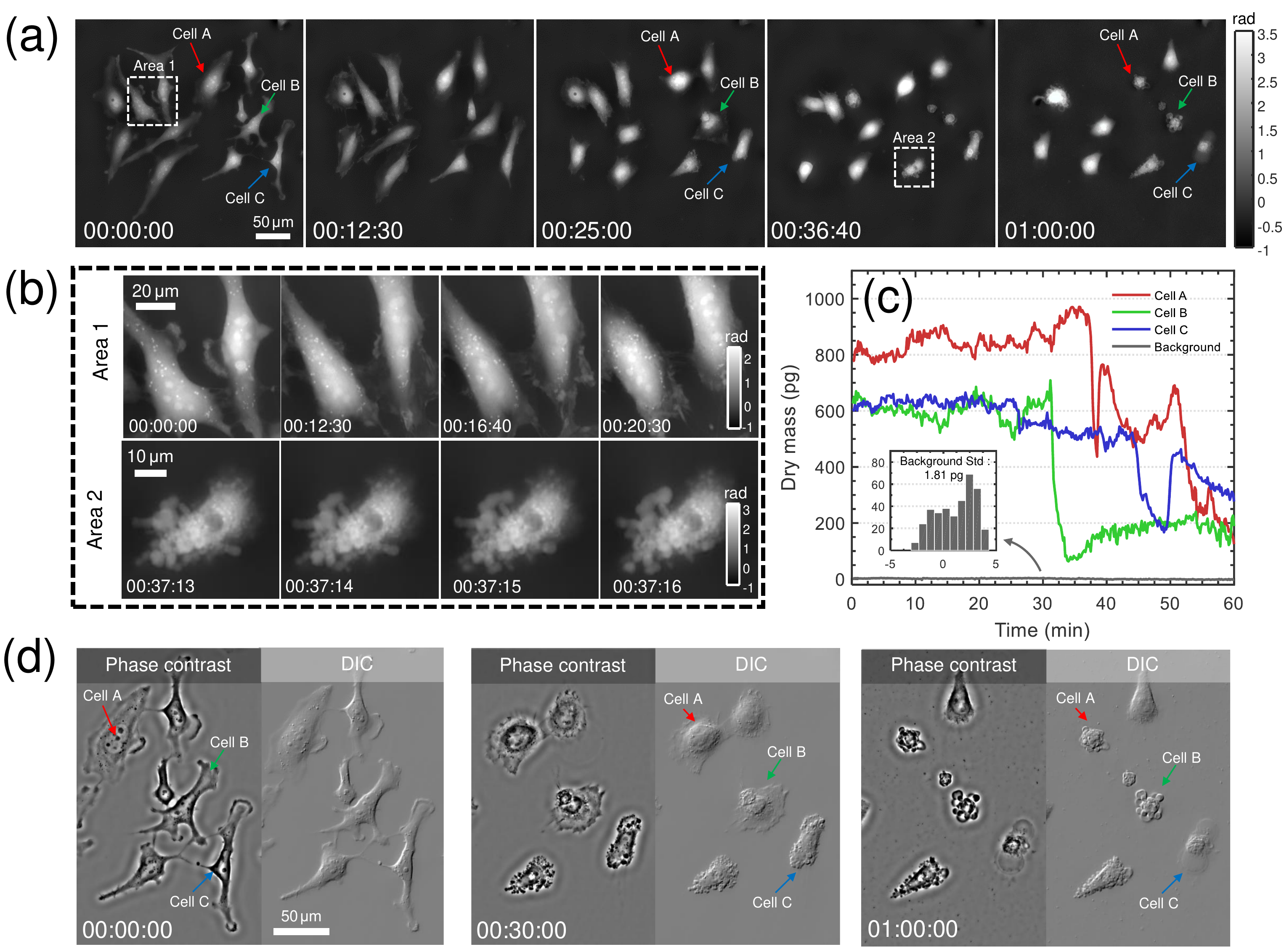}
\caption{Multi-modal computational imaging of apoptotic HeLa cells induced by paclitaxel. The phase is reconstructed every second from the beginning of the drug treatment. (a) 5 representative quantitative phase images spanning over 1 h. (b) Magnified views at different time points corresponding to regions of interest (Areas 1 and 2). (c) Dry mass changes of three labeled cells during apoptosis. (d) Simulated phase contrast and DIC images at three time points.}
\label{fig15}
\end{figure}

\noindent
Finally, AI-TIE is served as a multi-modal imaging tool for the visualization of morphology changes of apoptotic HeLa cells induced by paclitaxel. Paclitaxel (Taxol) is an antitumor drug that is remarkably effective against advanced ovarian carcinoma and many other types of tumors. Figure \ref{fig15}(a) shows five representative quantitative phase images over the period of 1 h from the beginning of drug treatment. During this time, most Hela cells contracted, rounded up, and/or disintegrated into small structures. Since the cell death process, including cell shrinkage, rounding up, and breaking down into fragments is highly dynamic and ever changing, fast acquisition speed is required to avoid motion blur. In AI-TIE, the high-quality phase can be recovered with only 3 images, which permits study of dynamic subcellular process with a temporal resolution of one second (1 frame/s). Figure \ref{fig15}(b) shows 4 reconstructed phases corresponding to two boxed areas of interest in Fig. \ref{fig15}(a) with finer time intervals (note that the frame interval of phase snapshots in Area 2 is only 1 s). The corresponding time-evolution video at this finer temporal resolution over the whole period of 1 h is provided in \textbf{Supplementary Video 3}. Such acquisition speed allows subcellular dynamics of Hela cell, such as moving vesicles, and shrinking lamellipodium being clearly observed without any motion-induced blurring or artifacts. From the vast amount of morphological parameters stored in each phase image, we chose here to analyze the change in dry mass of three individual cells (Cells A, B, C, labeled by the red, green, blue arrows) over time. It can be observed that the apoptosis of Hela cells is connected with the significant loss of dry mass during the cell shrinkage and fragmentation. The rebound in dry mass during late stage of apoptosis is mainly correlated with the formation of blebs and apoptotic bodies. Benefitting from the low-noise phase reconstruction provided by annular illumination TIE, the background dry mass fluctuations (by performing 360 independent measurements on 50-by-50 non-cellular region over the 1 h period) is only 1.81 pg (standard deviation), which is negligible compared to the active signals of the cells.

For morphological analysis of apoptotic cells, phase contrast and DIC images are created from quantitative phase maps, which are shown in Fig. \ref{fig15}(d) with a partial field of view (see \textbf{Supplementary Video 3} for corresponding time-evolution video). These images are particular useful for biologists who are used to observe samples with these well-established techniques. The phase contrast images show improved overall contrast of cell organelles, the DIC images highlight high-spatial-frequency subcellular details, while the quantitative phase images directly show the optical thickness of the object, facilitating cell segmentation and dry mass quantification. The major hallmarks of apoptosis, such as nucleus condensation, granulation of the cytoplasm, blebbing, fragment of the cell, and formation of apoptotic bodies are clearly visualized by those complementary three imaging modalities with both spatial resolution and low noise. It should be also noted that with short camera exposure time (50 ms), the current imaging speed  (1 frame/s) is ultimately limited by the $z$-scanning mechanics. There is still plenty of room for speed improvement (to video rate or even camera frame-rate limited speed), if a spatial light modulator \cite{26,34} or an electrically tunable lenses \cite{33,92} is used to replace the scanning mechanics.

\section*{Discussion}

\noindent
In this work, the effect of source distribution on the formation of phase information in a partially coherent microscope has been explored, and it is shown that the use of annular illumination matching the objective NA allows for high-quality, low-noise phase reconstruction with a lateral resolution close to the microscope incoherent diffraction limit. Compared with TIE phase imaging using a conventional circular illumination aperture and multiple defocus distances, AI-TIE achieves comparable noise-robustness, avoids the need for additional data acquisition, and more importantly, provides significant resolution improvement over its circular alternatives. This new approach enables us to demonstrate for the first time that the incoherent diffraction limited TIE quantitative phase imaging of unlabeled cells, achieving a lateral resolution of 208 nm, corresponding to an effective NA of 2.66 by incorporating high-NA oil-immersion annular illumination as well as high-NA oil-immersion objective detection. We also demonstrated the effectiveness of this approach by time-lapse imaging of \emph{in vitro} Hela cells. The high-resolution phase information clearly highlights cellular morphological and local biomolecule concentration changes during cells mitosis and apoptosis. Moreover, AI-TIE operates with a low-intensity illumination and very short exposure time. The lose-dose, high-NA annular illumination offers a high-resolution, non-invasive, and non-phototoxic means of quantifying biological behavior and dynamic variations over time. Since this technique does not require temporal coherent illumination, it is compatible with conventional brightfield microscope (halogen lamp, K\"ohler illumination), and can be easily combined with fluorescence techniques to gain molecular specificity and thus provide wider window to investigate biological process \cite{93,94}.

The theoretical analysis and experimental results suggest reshaping of illumination source provides new possibilities to push the resolution limit and improve low frequency performance of TIE imaging. However, due to the complicated form of partially coherent WOTF, solving for an optimum source pattern is quite challenging. The current choice of annular aperture was empirically designed based on intuitive criteria related to the shape of WOTF. Enabling the use of a more elaborate criterion (merit function) for evaluating the “goodness” of an aperture, and optimizing the aperture based on optimization algorithms are interesting directions for future work.

Recovering depth-resolved 3D phase information with the transport-of-intensity approach for thick phase objects is another important direction that requires further investigation. For imaging thick object, the 2D WOTF should be extended to 3D \cite{49,51,63}, and the annular illumination is expected to better frequency coverage and response for phase than conventional circular illumination in the 3D Fourier space, resulting in resolution enhancement both lateral and axial directions. The combination of annular illumination and focus-scanning may provide the possibility of achieving the same 3D resolution as the one obtained using diffraction tomographic holography \cite{19,95}, but significantly simplifies the setup.

\bibliography{RRef}

\section*{Acknowledgements}

This work was supported by the National Natural Science Fund of China (61505081, 111574152), Final Assembly ‘13th Five-Year Plan’ Advanced Research Project of China (30102070102), ‘Six Talent Peaks’ project of Jiangsu Province, China (2015-DZXX-009), ‘333 Engineering’ Research Project of Jiangsu Province, China (BRA2016407, BRA2015294), Fundamental Research Funds for the Central Universities (30917011204, 30916011322). C. Zuo thanks the support of the ‘Zijin Star’ program of Nanjing University of Science and Technology.

\section*{Author contributions statement}

C.Z. proposed the idea. C.Z. and J.S. developed the theoretical description of the method and performed WOTF calculations. C.Z., J.L. and J.Z. performed experiments. C.Z. and J.S. analyzed the data. C.Z., Q.C. and A.A. supervised the research. All authors contributed to writing the manuscript.

\section*{Additional information}

\textbf{Competing financial interests:} The authors declare no competing financial interests.

\end{document}